\documentclass[preprint,12pt]{elsarticle}

\usepackage{amssymb}
\usepackage{color}

\journal{J. Alloys Compd.}

\begin{document}

\begin{frontmatter}



\title{Enhanced $T_\mathrm{c}$ in eutectic high-entropy alloy superconductors Hf-Nb-Sc-Ti-Zr}


\author[inst1]{Issei Kubo}

\affiliation[inst1]{organization={Department of Electrical Engineering, Faculty of Engineering, Fukuoka Institute of Technology},%Department and Organization
            addressline={3-30-1 Wajiro-higashi, Higashi-ku}, 
            city={Fukuoka},
            postcode={811-0295},
            country={Japan}}

\author[inst2]{Yuto Watanabe}

\affiliation[inst2]{organization={Department of Physics, Tokyo Metropolitan University},%Department and Organization
            city={Hachioji},
            postcode={192-0397}, 
            country={Japan}}

\author[inst1]{Shuma Kawashima}

\author[inst1]{Tomohiro Miyaji}

\author[inst2]{Yoshikazu Mizuguchi}

\author[inst3]{Terukazu Nishizaki}

\affiliation[inst3]{organization={Department of Electrical Engineering, Faculty of Science and Engineering, Kyushu Sangyo University},%Department and Organization
            addressline={2-3-1 Matsukadai, Higashi-ku}, 
            city={Fukuoka},
            postcode={813-8503},
            country={Japan}}

\author[inst1]{Jiro Kitagawa}

\begin{abstract}
The present investigation into the superconducting properties of eutectic high-entropy alloy (HEA) Hf-Nb-Sc-Ti-Zr systems reveals an enhanced superconducting critical temperature ($T_\mathrm{c}$) in body-centered cubic (bcc) phases compared to typical quinary bcc HEAs. 
In Hf$_{10}$Nb$_{25}$Sc$_{25}$Ti$_{20}$Zr$_{20}$, Hf$_{5}$Nb$_{45}$Sc$_{20}$Ti$_{15}$Zr$_{15}$, and Hf$_{5}$Nb$_{45}$Sc$_{10}$Ti$_{5}$Zr$_{35}$ systems, which span a broad range of valence electron concentration per atom, lattice strain and the presence of partial or absent eutectic phases are characteristic features at lower annealing temperatures. 
The eutectic regions expand rapidly following annealing at 600 \hspace{1mm}$^{\circ}$C in all systems. 
The $T_\mathrm{c}$ of each system increases markedly with rising annealing temperatures from 400 \hspace{1mm}$^{\circ}$C to 600 \hspace{1mm}$^{\circ}$C, reaching a maximum value of 9.93 K in the Hf$_{5}$Nb$_{45}$Sc$_{10}$Ti$_{5}$Zr$_{35}$ sample annealed at 800 \hspace{1mm}$^{\circ}$C. 
Nearly all samples can be classified as strong-coupling superconductors. 
The sample annealed at 500 \hspace{1mm}$^{\circ}$C in the Hf$_{5}$Nb$_{45}$Sc$_{10}$Ti$_{5}$Zr$_{35}$ system exhibits a critical current density ($J_\mathrm{c}$) exceeding the practical threshold of 10$^{5}$ A/cm$^{2}$ up to approximately 4 T at 4.2 K and 6 T at 2 K. 
The elevated $J_\mathrm{c}$ is attributed to significant lattice strain and phase instability. 
The underlying mechanism for the enhanced $T_\mathrm{c}$ in Hf-Nb-Sc-Ti-Zr systems is examined through specific heat data analysis, suggesting that the expansion of the eutectic regions induced by thermal annealing plays a pivotal role.
\end{abstract}


%\item 
%\item 
%\item 
%\item 


\end{frontmatter}


\section{Introduction}
High-entropy alloys (HEAs) challenge traditional alloy design paradigms, wherein a few minor elements are incorporated into one or two primary constituents to enhance specific properties. 
The foundational principle of HEA design is the formation of solid solution phases stabilized by high configurational entropy\cite{Cantor:MSEA2004,Yeh:AEM2004,Gao:book,Cantor:book}. 
Consequently, HEAs represent an emerging class of multicomponent alloys comprising four or more principal elements\cite{Yan:MMTA2021}. 
These alloys exhibit various functionalities, including exceptional mechanical performance at elevated temperatures, superior corrosion resistance, wear-resistant hard coatings, and soft ferromagnetism, underscoring their growing importance in materials science\cite{Li:PMS2021,Fu:JMST2021,Li:AEM2019,Chaudhary:MT2021,Kitagawa:JMMM2022}. 
A particularly notable feature of HEAs—rarely observed in conventional alloys—is their multifunctionalities, such as the simultaneous realization of high strength and ductility or the coexistence of soft ferromagnetism and corrosion resistance\cite{Han:NRM2024,Lu:SM2020,Duan:SCM2023}. 
The demand for such advanced materials is increasing across critical sectors, including aerospace, nuclear energy, and defense.

Superconductivity is a key functional property of metallic alloys, and research into HEA superconductors has advanced significantly over the past decade\cite{Kozelj:PRL2014,Sun:PRM2019,Kitagawa:Metals2020,Kitagawa:book,Zeng:NPGAM2024,Kitagawa:EPJB2025}. 
Notably, the discovery of robust superconductivity under extreme conditions—such as high pressure and ion irradiation—and the observation of high critical current density ($J_\mathrm{c}$) in HEA thin films are considered groundbreaking developments\cite{Guo:PNAC2017,Kasem:SciRep2022,Jung:NC2022}. 
Multifunctionality is also anticipated in HEA superconductors, for instance, as superconducting wires suitable for aerospace or nuclear fusion applications. 
With a focus on body-centered cubic (bcc) alloy superconductors, conventional binary or ternary $d$-electron alloys exhibit the well-known Matthias rule, which correlates the superconducting critical temperature ($T_\mathrm{c}$) with valence electron concentration per atom (VEC), showing broad maxima at VEC values near 4.6 and 6.6\cite{Matthias:PR1955}. 
This rule has also been examined in numerous quinary bcc HEA superconductors within the VEC range of 4.1–5.2\cite{Hattori:JAMS2023}. 
Although quinary HEAs tend to exhibit Matthias rule-like behavior, their $T_\mathrm{c}$ values are significantly suppressed compared to conventional bcc alloys with equivalent VEC, likely due to the pronounced structural disorder inherent to HEAs.

We have recently reported the eutectic HEA superconductor NbScTiZr, which exhibits a distinctive annealing temperature dependence of $T_\mathrm{c}$ and an intriguing correlation between microstructure and $J_\mathrm{c}$\cite{Seki:JSNM2023,Kitagawa:MTC2024}. 
The eutectic structure of NbScTiZr comprises bcc and hexagonal close-packed (hcp) phases. 
Pioneering work by Krnel et al. revealed that the bcc phase undergoes a superconducting transition in the as-cast state\cite{Krnel:Materials2022}.
Detailed investigations by our research group into the effects of annealing on the superconducting properties of NbScTiZr demonstrated a marked enhancement of $T_\mathrm{c}$ from 7.9 K in the as-cast state to 9 K in the sample annealed at 800 \hspace{1mm}$^{\circ}$C\cite{Seki:JSNM2023,Kitagawa:MTC2024}. 
Structural characterization indicates the presence of lattice strain at lower annealing temperatures, corroborated by a reduction in the lattice parameter. 
The as-cast sample exhibits a fine eutectic lamellar-like structure with a thickness of approximately 70 nm. 
While this fine structure is preserved in the sample annealed at 400 \hspace{1mm}$^{\circ}$C, further increases in annealing temperature lead to significant grain coarsening.
An additional key feature of NbScTiZr is its elevated $J_\mathrm{c}$; the sample annealed at 400 \hspace{1mm}$^{\circ}$C achieves peak performance, exceeding the practical threshold of 10$^{5}$ A/cm$^{2}$ up to approximately 3.5 T at 4.2 K and 5.5 T at 2 K\cite{Kitagawa:EPJB2025}.
The high performance is attributed to the combined effects of pronounced lattice strain and the fine eutectic microstructure.

Notably, NbScTiZr samples, from the as-cast to 800 \hspace{1mm}$^{\circ}$C annealed states, exhibit a significant deviation from the Matthias rule-like behavior observed in many bcc HEAs; specifically, $T_\mathrm{c}$ is enhanced\cite{Seki:JSNM2023,Kitagawa:MTC2024}. 
According to BCS theory, $T_\mathrm{c}$ depends on two principal factors: the density of states at the Fermi level ($E_\mathrm{F}$), denoted as $D(E_\mathrm{F})$, and the electron-phonon interaction. 
The VEC reflects $D(E_\mathrm{F})$, and the Matthias rule is predicated on the assumption that $D(E_\mathrm{F})$ primarily governs $T_\mathrm{c}$\cite{Matthias:PR1955}. 
Consequently, the deviation observed in NbScTiZr implies that the eutectic structure may significantly influence the electron-phonon interaction. 
To evaluate the universality of this deviation from the typical trend observed in bcc HEAs, it is desirable to investigate quinary HEAs with eutectic microstructures, as many bcc superconducting HEAs are quinary in composition. 
Moreover, we aim to explore the relationship between microstructure and $J_\mathrm{c}$ in quinary eutectic HEAs.

In this study, we incorporated the element Hf into the NbScTiZr matrix, preparing the eutectic alloy systems Hf$_{10}$Nb$_{25}$Sc$_{25}$Ti$_{20}$Zr$_{20}$, Hf$_{5}$Nb$_{45}$Sc$_{20}$Ti$_{15}$Zr$_{15}$, and Hf$_{5}$Nb$_{45}$Sc$_{10}$Ti$_{5}$Zr$_{35}$, which span a wide range of VEC values. 
We systematically investigated the structural, metallurgical, and superconducting properties of these systems under varying annealing conditions. 
The comparison of superconducting properties among quinary bcc HEA superconductors was discussed through the VEC dependence of $T_\mathrm{c}$ and specific heat data analyses.

\section{Materials and Methods}
The as-cast polycrystalline samples were synthesized using a homemade arc furnace under an argon atmosphere. 
The constituent elements were Hf chips (Mitsuwa Chemicals, 99.6 \hspace{1mm}\%), Nb wire (Nilaco, 99.9 \hspace{1mm}\%), Sc chips (Kojundo Chemical Laboratory, 99 \hspace{1mm}\%), Ti wire (Nilaco, 99.9 \hspace{1mm}\%), and Zr wire (Nilaco, 99.5 \hspace{1mm}\%). 
Elemental mixtures with atomic ratios of Hf:Nb:Sc:Ti:Zr = 10:25:25:20:20, 5:45:20:15:15, or 5:45:10:5:35 were repeatedly melted with intermediate flipping to ensure homogeneity, followed by quenching on a water-cooled copper hearth. 
The nominal compositions were selected to span a broad range of VEC. 
The as-cast samples were sealed in evacuated quartz tubes and subjected to annealing at 400 \hspace{1mm}$^{\circ}$C, 500 \hspace{1mm}$^{\circ}$C, 600 \hspace{1mm}$^{\circ}$C, or 800 \hspace{1mm}$^{\circ}$C for four days using an electric furnace. 
The annealing temperatures of 400 \hspace{1mm}$^{\circ}$C, 600 \hspace{1mm}$^{\circ}$C, and 800 \hspace{1mm}$^{\circ}$C were adopted in accordance with our previous study\cite{Kitagawa:MTC2024}, while an additional condition of 500 \hspace{1mm}$^{\circ}$C was introduced to examine the detailed annealing-temperature dependence of the lattice parameter of the bcc phase.
No mechanical deformation (e.g., rolling) was applied during sample preparation.

Room-temperature X-ray diffraction (XRD) patterns were collected using a Shimadzu XRD-7000L diffractometer equipped with Cu-K$\alpha$ radiation in Bragg–Brentano geometry. 
Due to the difficulty in producing fine powders, thin slabs sectioned from the bulk samples were used. 
Microstructural observations were conducted via field-emission scanning electron microscopy (FE-SEM, JEOL JSM-7100F). 
Elemental compositions were analyzed using an energy-dispersive X-ray (EDX) spectroscopy system integrated with the FE-SEM.
The microstructures were observed on the polished surfaces.
For the microstructural observation and EDX analysis, the sample surfaces were sequentially polished using silicon carbide papers (\#240, \#400, \#600, and \#1000 mesh), alumina pastes (5, 1, and 0.1 $\mu$m), and colloidal silica (0.04 $\mu$m).

The temperature dependence of the dc magnetization $M$($T$) and the isothermal magnetization curves were measured using a SQUID magnetometer (MPMS3, Quantum Design). 
For the $M$($T$) measurements, the sample was first cooled below $T_\mathrm{c}$ in zero field, after which an external magnetic field of 0.4 mT was applied, and $M$ was measured during warming (zero-field-cooled warming protocol). 
Subsequently, the sample was cooled again below $T_\mathrm{c}$ under a 0.4 mT field, and $M$ was measured during warming (field-cooled warming protocol). 
The isothermal magnetization curves were obtained under external magnetic fields sweeping between -7 T and 7 T.
Temperature-dependent electrical resistivity $\rho$($T$) measurements under magnetic fields of 0, 0.5, 1, 2, 3, 4, 5, 6, 7, 8, and 9 T were performed using the standard four-probe technique with a PPMS system (Quantum Design). 
Specific heat $C_\mathrm{p}$ was measured using the thermal-relaxation method, also employing the PPMS apparatus.

\section{Results and Discussion}
\subsection{Structural and Metallographic Characterizations}
\begin{figure}
\begin{center}
\includegraphics[width=1\linewidth]{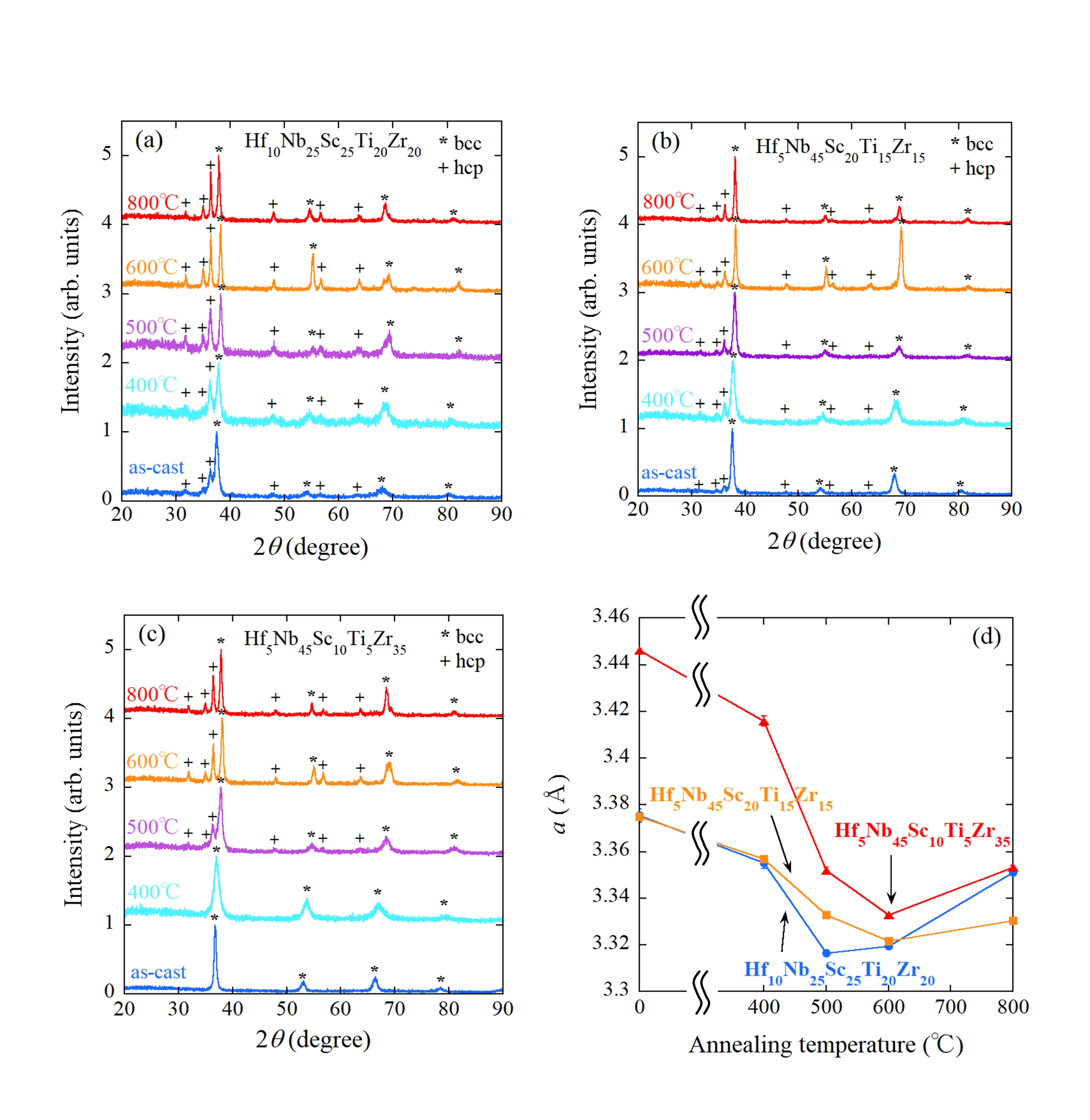}
\caption{\label{fig1}XRD patterns of (a) Hf$_{10}$Nb$_{25}$Sc$_{25}$Ti$_{20}$Zr$_{20}$, (b) Hf$_{5}$Nb$_{45}$Sc$_{20}$Ti$_{15}$Zr$_{15}$, and (c) Hf$_{5}$Nb$_{45}$Sc$_{10}$Ti$_{5}$Zr$_{35}$ samples prepared under various annealing conditions. The origin of each XRD pattern is vertically offset for the clarity. (d) Dependence of lattice parameter on annealing temperature for Hf$_{10}$Nb$_{25}$Sc$_{25}$Ti$_{20}$Zr$_{20}$, Hf$_{5}$Nb$_{45}$Sc$_{20}$Ti$_{15}$Zr$_{15}$, and Hf$_{5}$Nb$_{45}$Sc$_{10}$Ti$_{5}$Zr$_{35}$ systems.}
\end{center}
\end{figure}

Figures \ref{fig1}(a)-(c) present the XRD patterns of all investigated samples. 
In the Hf$_{10}$Nb$_{25}$Sc$_{25}$Ti$_{20}$Zr$_{20}$ and Hf$_{5}$Nb$_{45}$Sc$_{20}$Ti$_{15}$Zr$_{15}$ alloy systems, the Bragg reflection peaks observed in the as-cast samples can be attributed to either the bcc (*) or hcp (+) phases. 
In both systems, the intensity of the hcp phase increases progressively from the as-cast state to the 600 \hspace{1mm}$^{\circ}$C annealed condition. 
Concomitant with this evolution, the Bragg peaks of the bcc phase systematically shift to higher angles, indicating a significant reduction in the lattice parameter (see Figs.\hspace{1 mm}\ref{fig1}(a) and (b)).
In contrast, the XRD pattern of the as-cast Hf$_{5}$Nb$_{45}$Sc$_{10}$Ti$_{5}$Zr$_{35}$ exhibits a single-phase bcc structure, which is preserved in the sample annealed at 400 \hspace{1mm}$^{\circ}$C, although with broadened diffraction peaks. 
Further elevation of the annealing temperature results in the emergence of the hcp phase alongside the bcc phase, accompanied by a pronounced shift of the bcc Bragg reflections toward higher angles. 
Upon assigning Miller indices to the observed Bragg peaks, the lattice parameters of both phases were calculated using the least-squares method. 
Table \ref{tab1} summarizes the resulting lattice parameters for all phases across the different systems.
The annealing-temperature dependence of the bcc lattice parameters is plotted in Fig.\hspace{1mm}\ref{fig1}(d), where the as-cast state is treated as the 0 \hspace{1mm}$^{\circ}$C annealing. 
In each alloy system, the lattice parameter exhibits a sharp decrease above approximately 400 \hspace{1mm}$^{\circ}$C, reaching a minimum at 500-600 \hspace{1mm}$^{\circ}$C. 
This behavior mirrors that observed in the eutectic NbScTiZr system and is attributed to thermally induced lattice strain\cite{Seki:JSNM2023,Kitagawa:MTC2024}. 
It is also noteworthy that similar lattice parameter contraction upon thermal annealing has been reported in Ta$_{1/6}$Nb$_{2/6}$Hf$_{1/6}$Zr$_{1/6}$Ti$_{1/6}$ and (TaNb)$_{0.7}$(HfZrTi)$_{0.5}$\cite{Kim:JMST2024,Gao:APL2022}.

\begin{table}
\caption{Chemical compositions of bcc and hcp phases, lattice parameters, and VEC of prepared samples.}\label{tab1}%
\scriptsize
\begin{tabular}{cccc}
\hline
sample  & composition   &  lattice parameter (\AA) & VEC  \\
\hline
Hf$_{10}$Nb$_{25}$Sc$_{25}$Ti$_{20}$Zr$_{20}$ & & & \\
as-cast    & bcc:Hf$_{12.2(1)}$Nb$_{28.9(3)}$Sc$_{20.2(7)}$Ti$_{19.0(3)}$Zr$_{19.7(5)}$ & $a$=3.375(3) & 4.09  \\
           & hcp:Hf$_{8.8(2)}$Nb$_{19.3(4)}$Sc$_{33.3(4)}$Ti$_{18.3(3)}$Zr$_{20.3(2)}$ & $a$=3.251(1), $c$=5.133(2)  & 3.86   \\
400 \hspace{1mm}$^{\circ}$C & bcc:Hf$_{12.5(3)}$Nb$_{30.1(7)}$Sc$_{19.0(6)}$Ti$_{19.1(3)}$Zr$_{19.3(2)}$ & $a$=3.355(2) & 4.11  \\
           & hcp:Hf$_{9.4(2)}$Nb$_{20.7(5)}$Sc$_{31.1(5)}$Ti$_{18.6(2)}$Zr$_{20.1(3)}$ & $a$=3.250(1), $c$=5.132(2)  & 3.90   \\
500 \hspace{1mm}$^{\circ}$C & bcc:Hf$_{12.7(1)}$Nb$_{31.6(5)}$Sc$_{17.4(5)}$Ti$_{19.7(3)}$Zr$_{18.5(5)}$ & $a$=3.316(1) & 4.14  \\
           & hcp:Hf$_{9.3(9)}$Nb$_{20(1)}$Sc$_{31(1)}$Ti$_{19.0(9)}$Zr$_{20.2(5)}$ & $a$=3.246(1), $c$=5.124(2)  & 3.89   \\
600 \hspace{1mm}$^{\circ}$C & bcc:Hf$_{9.4(3)}$Nb$_{35.3(4)}$Sc$_{15.9(8)}$Ti$_{25.1(3)}$Zr$_{14.2(4)}$ & $a$=3.319(1) & 4.19  \\
           & hcp:Hf$_{11.3(5)}$Nb$_{20(1)}$Sc$_{29(1)}$Ti$_{17(1)}$Zr$_{23(1)}$ & $a$=3.242(1), $c$=5.117(2)  & 3.91   \\
800 \hspace{1mm}$^{\circ}$C & bcc:Hf$_{10.3(1)}$Nb$_{36(1)}$Sc$_{14(1)}$Ti$_{25.0(7)}$Zr$_{14.7(9)}$ & $a$=3.351(1) & 4.22  \\
           & hcp:Hf$_{11.5(1)}$Nb$_{17.2(3)}$Sc$_{32.8(3)}$Ti$_{14.7(3)}$Zr$_{23.8(2)}$ & $a$=3.246(1), $c$=5.117(2)  & 3.84   \\
Hf$_{5}$Nb$_{45}$Sc$_{20}$Ti$_{15}$Zr$_{15}$ & & & \\
as-cast    & bcc:Hf$_{5.7(1)}$Nb$_{53(1)}$Sc$_{11.7(9)}$Ti$_{16.0(6)}$Zr$_{13.1(4)}$ & $a$=3.375(2) & 4.42  \\
           & hcp:Hf$_{4.9(3)}$Nb$_{21(1)}$Sc$_{44(1)}$Ti$_{10.6(4)}$Zr$_{20(1)}$ & $a$=3.268(1), $c$=5.166(2)  & 3.77   \\
400 \hspace{1mm}$^{\circ}$C & bcc:Hf$_{5.4(1)}$Nb$_{57(1)}$Sc$_{10.1(5)}$Ti$_{15.5(8)}$Zr$_{12.0(4)}$ & $a$=3.357(1) & 4.47  \\
           & hcp:Hf$_{4.4(1)}$Nb$_{18(1)}$Sc$_{51(1)}$Ti$_{8.5(7)}$Zr$_{17.9(7)}$ & $a$=3.264(1), $c$=5.159(1)  & 3.67   \\
500 \hspace{1mm}$^{\circ}$C & bcc:Hf$_{5.8(1)}$Nb$_{59.4(5)}$Sc$_{8.2(3)}$Ti$_{14.8(2)}$Zr$_{11.7(1)}$ & $a$=3.333(1) & 4.51  \\
           & hcp:Hf$_{4.5(2)}$Nb$_{19(1)}$Sc$_{49(1)}$Ti$_{9.3(6)}$Zr$_{18.2(2)}$ & $a$=3.270(2), $c$=5.151(4)  & 3.70   \\
600 \hspace{1mm}$^{\circ}$C & bcc:Hf$_{5.7(3)}$Nb$_{52.9(1)}$Sc$_{12.1(1)}$Ti$_{16.5(3)}$Zr$_{12.8(5)}$ & $a$=3.322(1) & 4.41  \\
           & hcp:Hf$_{4.6(2)}$Nb$_{20(1)}$Sc$_{48(1)}$Ti$_{8.2(1)}$Zr$_{19.2(1)}$ & $a$=3.258(1), $c$=5.139(2)  & 3.72   \\
800 \hspace{1mm}$^{\circ}$C & bcc:Hf$_{5.1(1)}$Nb$_{56(1)}$Sc$_{10(1)}$Ti$_{18.0(5)}$Zr$_{10.6(6)}$ & $a$=3.330(1) & 4.46  \\
           & hcp:Hf$_{6.8(2)}$Nb$_{10(1)}$Sc$_{47.8(8)}$Ti$_{5.8(2)}$Zr$_{29.2(5)}$ & $a$=3.260(1), $c$=5.144(2)  & 3.62   \\
Hf$_{5}$Nb$_{45}$Sc$_{10}$Ti$_{5}$Zr$_{35}$ & & & \\
as-cast    & bcc (bright):Hf$_{5.9(1)}$Nb$_{50(1)}$Sc$_{7(1)}$Ti$_{4.7(2)}$Zr$_{32(1)}$ & $a$=3.446(1) & 4.43  \\
           & bcc (dark):Hf$_{5.4(2)}$Nb$_{38(1)}$Sc$_{13(1)}$Ti$_{5.1(4)}$Zr$_{38.8(9)}$ & $a$=3.446(1) & 4.25   \\
400 \hspace{1mm}$^{\circ}$C & bcc (bright):Hf$_{6.1(1)}$Nb$_{51.8(5)}$Sc$_{6.0(4)}$Ti$_{4.8(4)}$Zr$_{31.2(7)}$ & $a$=3.416(2) & 4.46  \\
           & bcc (dark):Hf$_{5.4(2)}$Nb$_{38(1)}$Sc$_{12.6(8)}$Ti$_{5.3(3)}$Zr$_{39(1)}$ & $a$=3.416(2) & 4.25   \\
500 \hspace{1mm}$^{\circ}$C & bcc (bright):Hf$_{5.9(3)}$Nb$_{48.3(4)}$Sc$_{7.9(1)}$Ti$_{4.8(3)}$Zr$_{33.2(3)}$ & $a$=3.352(1) & 4.40  \\
           & bcc (dark):Hf$_{5.6(2)}$Nb$_{39(1)}$Sc$_{12.0(5)}$Ti$_{5.2(4)}$Zr$_{37.8(9)}$ & $a$=3.352(1) & 4.27   \\
                & hcp: not identified due to fine microstructure & $a$=3.251(3), $c$=5.128(6) & - \\
600 \hspace{1mm}$^{\circ}$C & bcc (bright):Hf$_{6.2(3)}$Nb$_{55.8(1)}$Sc$_{5.7(2)}$Ti$_{4.0(2)}$Zr$_{28.4(2)}$ & $a$=3.333(1) & 4.50  \\
           & bcc (dark):Hf$_{5.5(2)}$Nb$_{36(1)}$Sc$_{14(1)}$Ti$_{5.0(3)}$Zr$_{39(1)}$ & $a$=3.333(1) & 4.22   \\
                & hcp: not identified due to fine microstructure & $a$=3.241(1), $c$=5.126(2) & - \\
800 \hspace{1mm}$^{\circ}$C & bcc:Hf$_{4.6(3)}$Nb$_{65(1)}$Sc$_{3(1)}$Ti$_{6.6(3)}$Zr$_{21(1)}$ & $a$=3.353(1) & 4.62  \\
           & hcp:Hf$_{6.5(2)}$Nb$_{14(1)}$Sc$_{23.3(1)}$Ti$_{2.7(1)}$Zr$_{54(1)}$ & $a$=3.239(1), $c$=5.135(2)  & 3.91   \\       
\hline
\end{tabular}
\end{table}

\begin{figure}
\begin{center}
\includegraphics[width=1\linewidth]{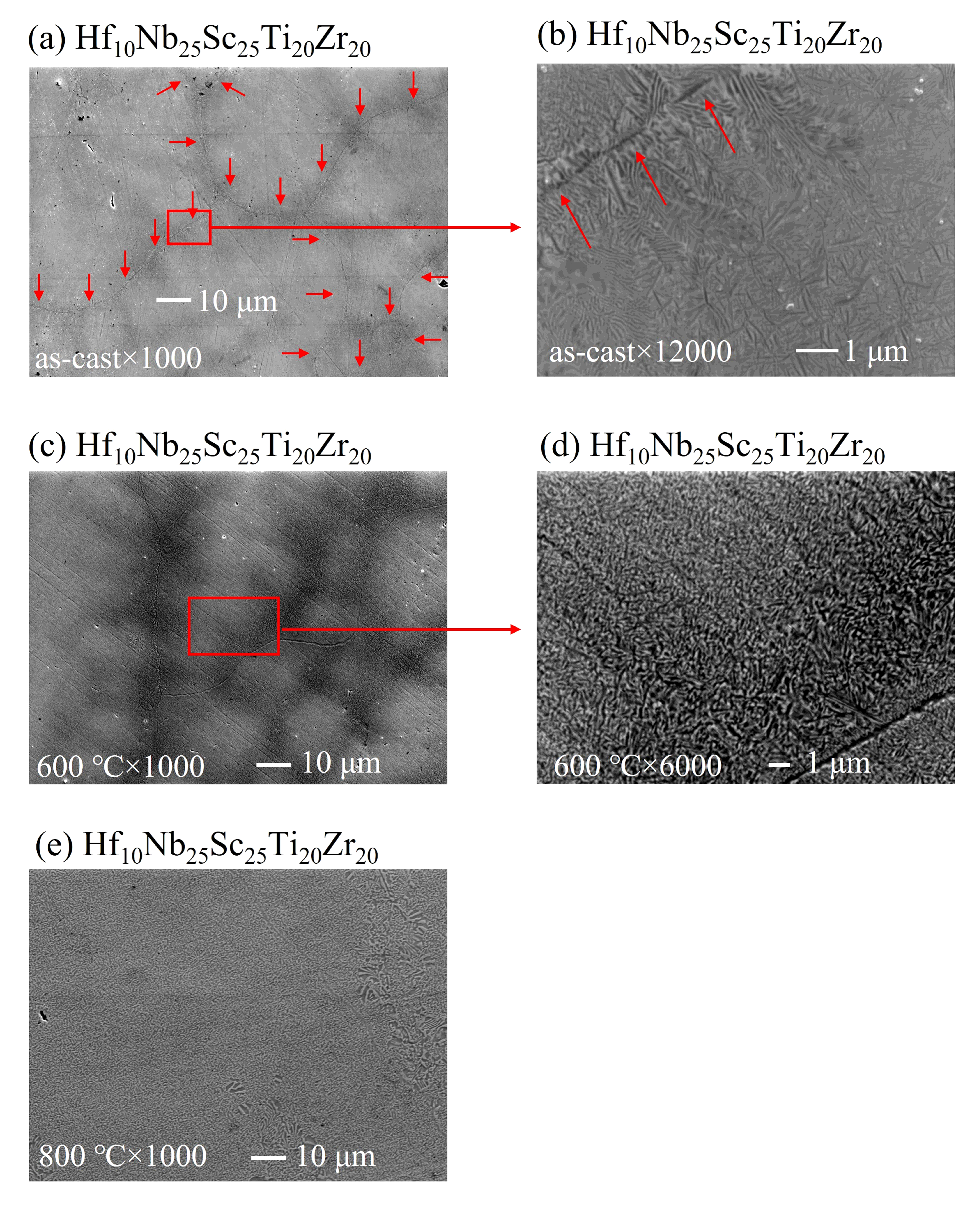}
\caption{\label{fig2}SEM images of Hf$_{10}$Nb$_{25}$Sc$_{25}$Ti$_{20}$Zr$_{20}$ for as-cast sample (a) and (b) and heat-treated samples at (c) and (d) 600 \hspace{1mm}$^{\circ}$C, and (e) 800 \hspace{1mm}$^{\circ}$C, respectively.}
\end{center}
\end{figure}

Figures \ref{fig2}(a)-(e) display SEM images of Hf$_{10}$Nb$_{25}$Sc$_{25}$Ti$_{20}$Zr$_{20}$ samples subjected to various annealing treatments. 
The chemical compositions determined by EDX for all samples are summarized in Table \ref{tab1}. 
A dendritic microstructure is observed in the as-cast sample (Fig.\hspace{1 mm}\ref{fig2}(a)). 
The VEC of the bright dendritic phase exceeds 4.0, whereas the interdendritic dark phase exhibits a VEC below 4.0 (see also Table \ref{tab1}). 
According to the phase stability study of alloys containing Sc and group 4 and 5 elements, the bcc and hcp phases are stabilized at VEC values above and below 4.0, respectively\cite{YM:JALCOM2022}. 
This VEC–structure relationship has also been confirmed in NbScTiZr\cite{Seki:JSNM2023}, supporting the interpretation that the bright and dark phases correspond to the bcc and hcp structures, respectively.
A magnified image (Fig.\hspace{1 mm}\ref{fig2}(b)) reveals a weave-like microstructure. 
Notably, a characteristic line, marked by red arrows, runs along the grain boundaries between the bcc and hcp phases and is accompanied by a lamellar-like structure. 
Annealing at 400 \hspace{1mm}$^{\circ}$C does not significantly alter the microstructure, although more pronounced lamellar features with spacings of approximately 80 nm partially emerge within grains in the 500 \hspace{1mm}$^{\circ}$C annealed sample (see Figs.\hspace{1mm}S1(a)-(e) in the Supplementary Material). 
At 600 \hspace{1mm}$^{\circ}$C, both bcc and hcp phases are filled with a eutectic microstructure, as revealed by the high-magnification image ($\times$ 6000, Fig.\hspace{1 mm}\ref{fig2}(d)). 
The lower-magnification image of the 800 \hspace{1mm}$^{\circ}$C annealed sample (Fig.\hspace{1 mm}\ref{fig2}(e)) also exhibits a developed eutectic structure and evidence of grain coarsening compared to the 600 \hspace{1mm}$^{\circ}$C sample. 
The VEC of the bcc phase increases systematically with annealing temperature, as listed in Table \ref{tab1}, which originates from a compositional shift toward a (Nb,Ti)-rich and Zr-poor state relative to the as-cast condition.

\begin{figure}
\begin{center}
\includegraphics[width=1\linewidth]{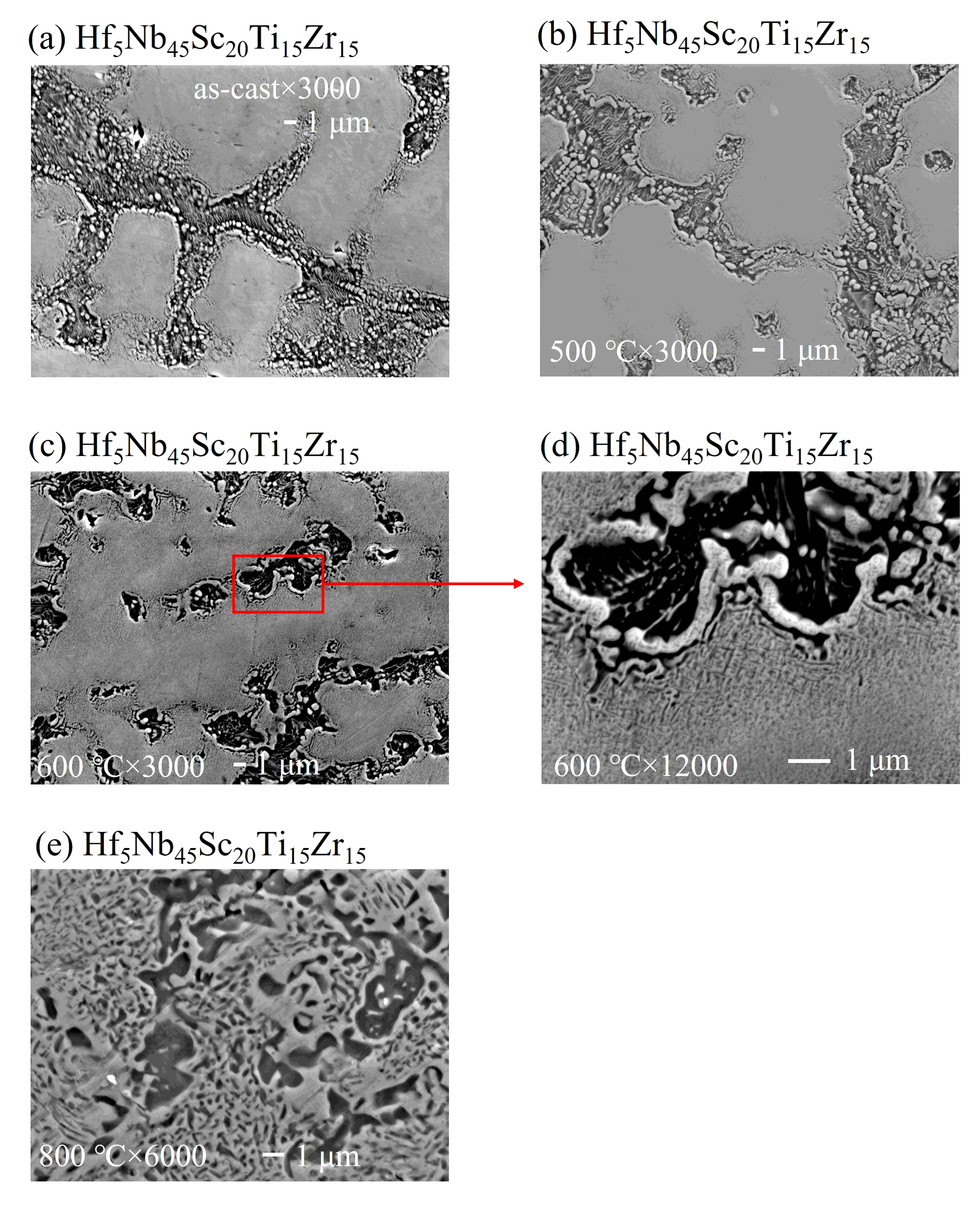}
\caption{\label{fig3}SEM images of Hf$_{5}$Nb$_{45}$Sc$_{20}$Ti$_{15}$Zr$_{15}$ for as-cast sample (a) and heat-treated samples at (b) 500 \hspace{1mm}$^{\circ}$C, (c) and (d) 600 \hspace{1mm}$^{\circ}$C, and (e) 800 \hspace{1mm}$^{\circ}$C, respectively.}
\end{center}
\end{figure}

In the SEM image of the as-cast Hf$_{5}$Nb$_{45}$Sc$_{20}$Ti$_{15}$Zr$_{15}$ sample, the large bright grains are assigned to the bcc phase based on the VEC–phase relationship (Fig.\hspace{1 mm}\ref{fig3}(a) and Table \ref{tab1}). 
A eutectic structure is discernible between the large bcc grains. 
This morphology is retained in the 400 \hspace{1mm}$^{\circ}$C and 500 \hspace{1mm}$^{\circ}$C annealed samples (Fig.\hspace{1mm}S2 in the Supplementary Material and Fig.\hspace{1 mm}\ref{fig3}(b)). 
Annealing at 600 \hspace{1mm}$^{\circ}$C induces a fine eutectic structure within the primary bcc phase (Figs.\hspace{1 mm}\ref{fig3}(c) and (d)). 
The grain size of the eutectic structure increases upon annealing at 800 \hspace{1mm}$^{\circ}$C, as shown in Fig.\hspace{1 mm}\ref{fig3}(e). 
The annealing effect on the VEC of the bcc phase in Hf$_{5}$Nb$_{45}$Sc$_{20}$Ti$_{15}$Zr$_{15}$ is negligible, as indicated in Table \ref{tab1}.

\begin{figure}
\begin{center}
\includegraphics[width=1\linewidth]{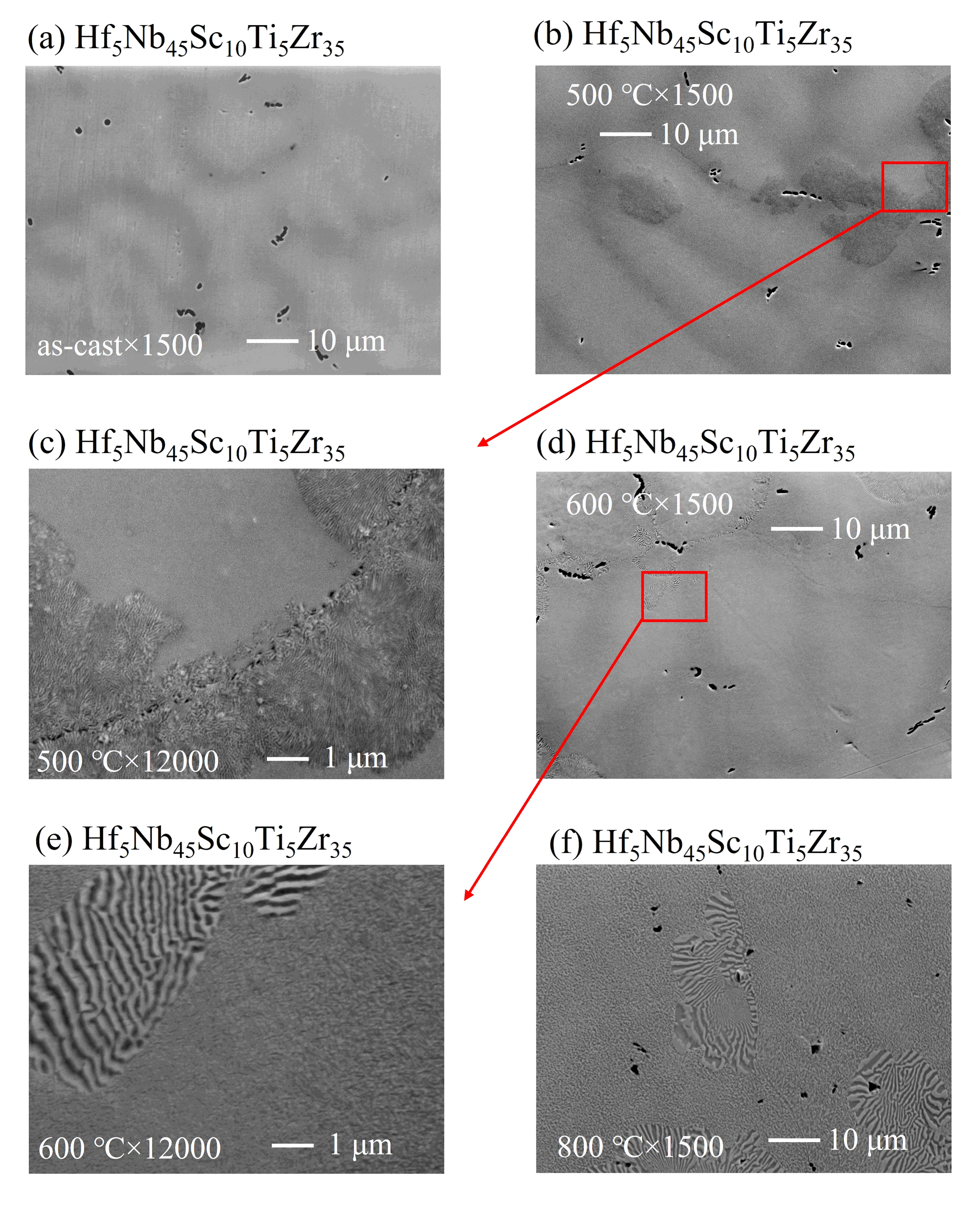}
\caption{\label{fig4}SEM images of Hf$_{5}$Nb$_{45}$Sc$_{10}$Ti$_{5}$Zr$_{35}$ for as-cast sample (a) and heat-treated samples at (b) and (c) 500 \hspace{1mm}$^{\circ}$C, (d) and (e) 600 \hspace{1mm}$^{\circ}$C, and (f) 800 \hspace{1mm}$^{\circ}$C, respectively.}
\end{center}
\end{figure}

Although the XRD pattern of the as-cast Hf$_{5}$Nb$_{45}$Sc$_{10}$Ti$_{5}$Zr$_{35}$ sample suggests a single-phase bcc structure, the corresponding SEM image reveals a phase-segregated microstructure ((Fig.\hspace{1 mm}\ref{fig4}(a)). 
Table \ref{tab1} lists the chemical compositions of both the bright and dark regions, with VEC values of 4.43 and 4.25, respectively—consistent with those stabilizing bcc HEAs\cite{YM:JALCOM2022}. 
Thus, the as-cast state is described as a dual bcc-phase structure with slightly differing VECs. 
The 400 \hspace{1mm}$^{\circ}$C annealed sample maintains a similar microstructure (see Fig.\hspace{1mm}S3 in the Supplementary Material).
In the 500 \hspace{1mm}$^{\circ}$C annealed sample, subtle microstructural evolution is observed (Figs.\hspace{1 mm}\ref{fig4}(b) and (c)), with very fine eutectic structures, approximately 40 nm in spacing, appearing in some dark regions. 
This microstructural change aligns with the emergence of hcp reflections in the XRD pattern (see also Fig.\hspace{1 mm}\ref{fig1}(c)). 
At 600 \hspace{1mm}$^{\circ}$C, fine eutectic structures also appear in the bright regions (Figs.\hspace{1 mm}\ref{fig4}(d) and (e)). 
Figure 4(f) illustrates the overall grain coarsening following 800 \hspace{1mm}$^{\circ}$C annealing.
Due to the extremely fine eutectic features, determining the precise chemical composition of the hcp phase in the 500 \hspace{1mm}$^{\circ}$C and 600 \hspace{1mm}$^{\circ}$C annealed samples was challenging. 
Therefore, the compositions of the bright and dark regions in Figs.\hspace{1 mm}\ref{fig4}(b) and (d) are reported in Table \ref{tab1}. 
While the VEC values of the bright and dark phases remain unchanged from the as-cast to the 500 \hspace{1mm}$^{\circ}$C annealed states, a systematic slight increase in the VEC of the bright phase is observed at 600 \hspace{1mm}$^{\circ}$C and 800 \hspace{1mm}$^{\circ}$C. 
This increase results from compositional changes trending toward Nb-rich and Zr-deficient compositions with increasing annealing temperature.

\subsection{Magnetization, electrical resistivity, and specific heat}
\begin{figure}
\begin{center}
\includegraphics[width=1\linewidth]{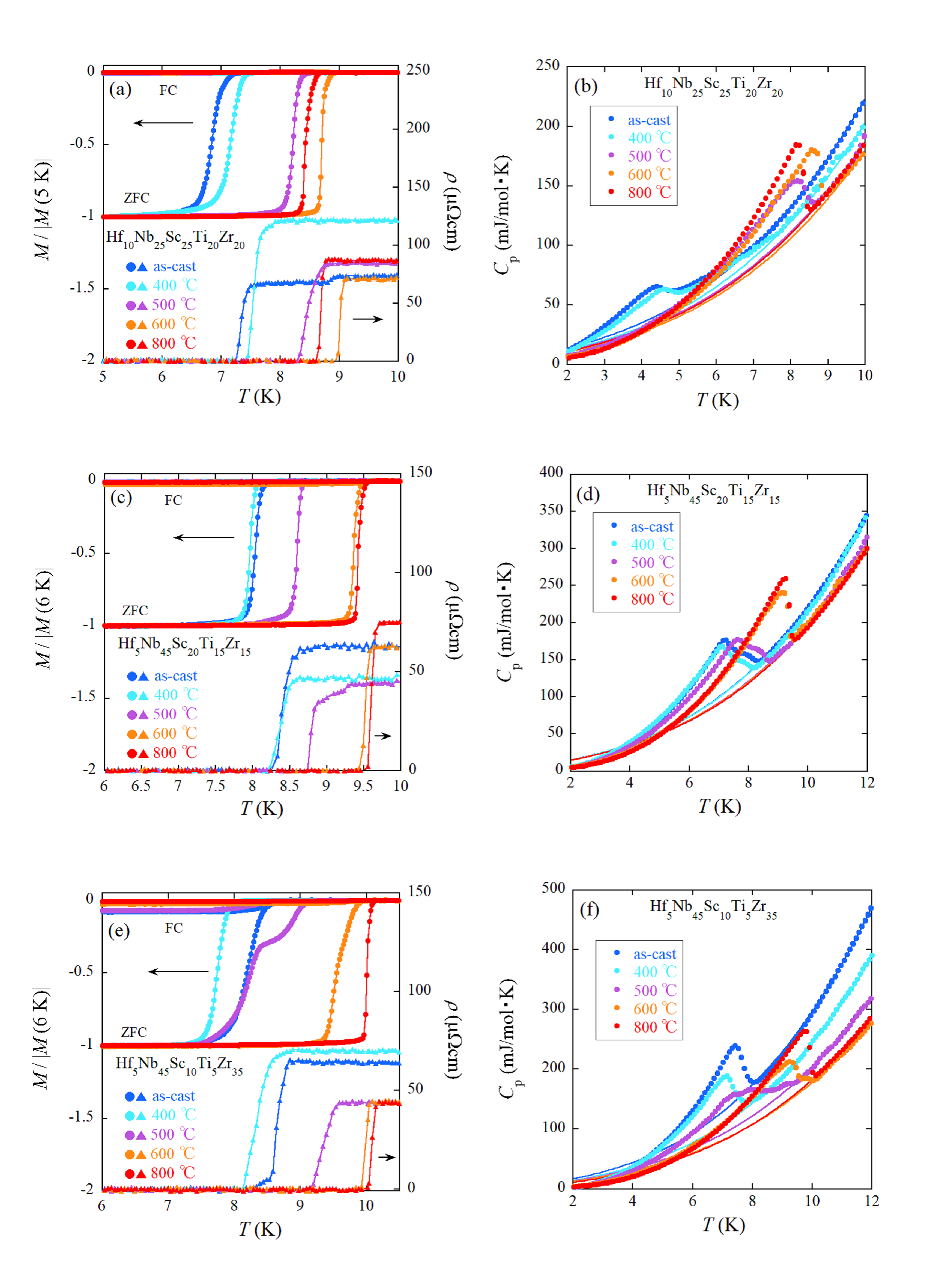}
\caption{\label{fig5} Temperature dependence of $M$ and $\rho$ for (a) Hf$_{10}$Nb$_{25}$Sc$_{25}$Ti$_{20}$Zr$_{20}$, (c) Hf$_{5}$Nb$_{45}$Sc$_{20}$Ti$_{15}$Zr$_{15}$, and (e) Hf$_{5}$Nb$_{45}$Sc$_{10}$Ti$_{5}$Zr$_{35}$ under various annealing conditions. Each $M$ value is normalized to its absolute value at 5 K or 6 K. Plots of $C_\mathrm{p}$ versus $T$ for (b) Hf$_{10}$Nb$_{25}$Sc$_{25}$Ti$_{20}$Zr$_{20}$, (d) Hf$_{5}$Nb$_{45}$Sc$_{20}$Ti$_{15}$Zr$_{15}$, and (f) Hf$_{5}$Nb$_{45}$Sc$_{10}$Ti$_{5}$Zr$_{35}$ under various annealing conditions. The solid line in each figure represents the fitting result using $\gamma T+\beta T^{3}$, where $\gamma$ is the electronic specific heat coefficient and $\beta T^{3}$ corresponds to the phonon part.}
\end{center}
\end{figure}

Figures \ref{fig5}(a), (c), and (e) summarize the temperature-dependent magnetization $M$($T$) and electrical resistivity $\rho$($T$) for all systems. 
$M$($T$) was measured under an external field $\mu_{0}H$ ($\mu_{0}$: the vacuum permeability) of 0.4 mT. 
In the figures, ZFC and FC refer to the zero-field-cooled and field-cooled measurement protocols.
$\rho$($T$) data were acquired under zero external magnetic field. 
We confirmed the presence of a diamagnetic signal and zero resistivity in each sample. 
Although the onset temperature of superconductivity in $\rho$($T$) tends to be higher than that observed in ZFC $M$($T$), Figs.\hspace{1mm}S4(a)-(c), which present ZFC $M$($T$) on an expanded scale in the Supplementary Material, demonstrate good agreement in the onset temperature between $\rho$($T$) and $M$($T$) for most samples. 
The specific heat $C_\mathrm{p}$ data for all samples are plotted as a function of temperature in Figs.\hspace{1 mm}\ref{fig5}(b), (d), and (f).

\begin{figure}
\begin{center}
\includegraphics[width=1.1\linewidth]{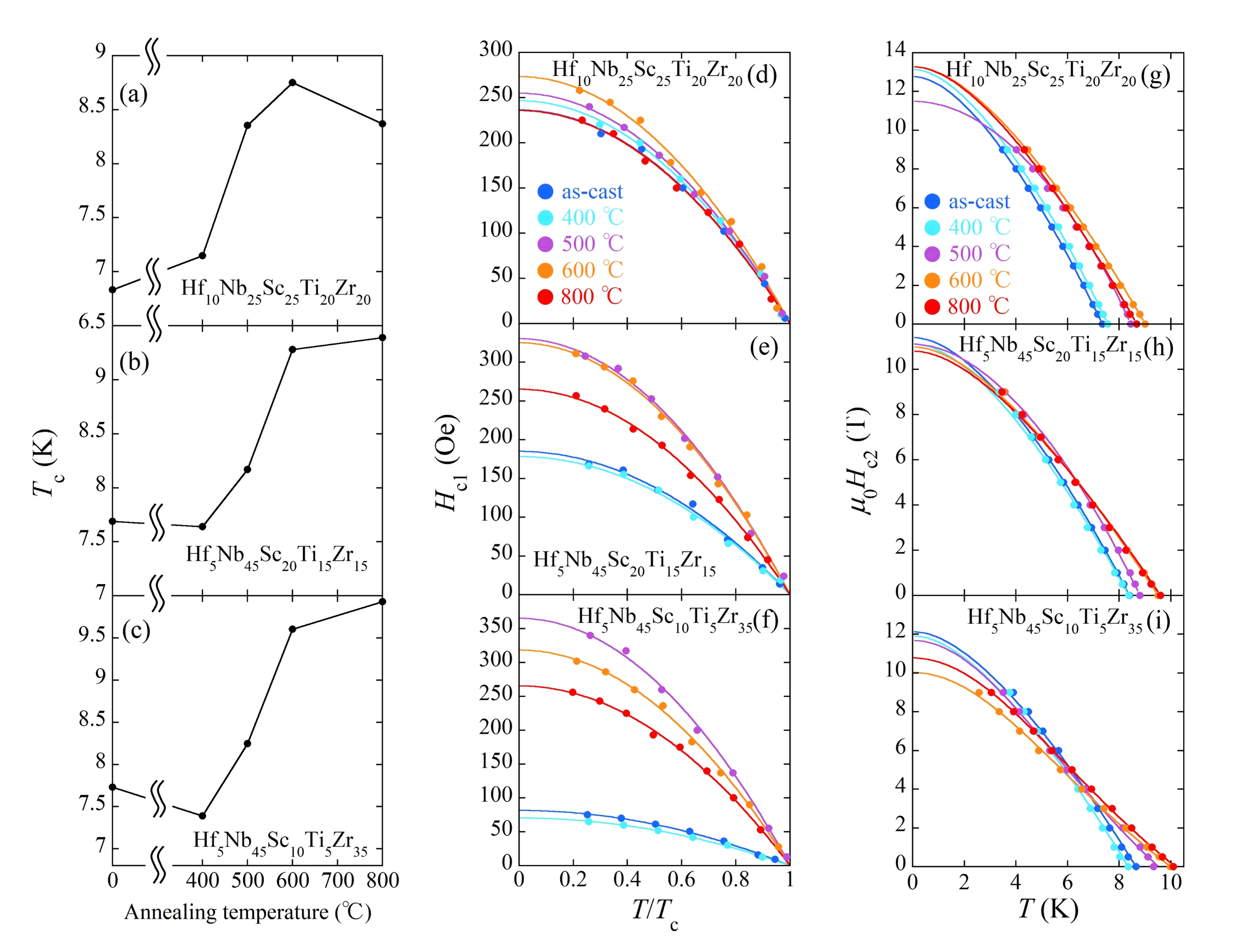}
\caption{\label{fig6} (a)-(c) Annealing temperature dependence of $T_\mathrm{c}$ for all alloy systems. $T_\mathrm{c}$ is defined as the midpoint of the specific heat jump, except for the as-cast and 400 \hspace{1mm}$^{\circ}$C annealed samples of Hf$_{10}$Nb$_{25}$Sc$_{25}$Ti$_{20}$Zr$_{20}$ and the 500 \hspace{1mm}$^{\circ}$C annealed sample of Hf$_{5}$Nb$_{45}$Sc$_{10}$Ti$_{5}$Zr$_{35}$. For these three samples, $T_\mathrm{c}$ is determined using the midpoint of the diamagnetic transition. Temperature dependence of (d)-(f) lower critical field and  (g)-(i) upper critical field for all alloy systems. In (d)-(f), the solid curves represent the fitting results using Eq.\hspace{1mm}(\ref{eq:hc1}). In (g)-(i), the solid curves correspond to the fitting results using Eq.\hspace{1mm}(\ref{eq:hc2}) or $H_\mathrm{c2}(T)=H_\mathrm{c2}(0)(1-(T/T_\mathrm{c})^{2})/(1+(T/T_\mathrm{c})^{2})$.}
\end{center}
\end{figure}

As shown in Fig.\hspace{1 mm}\ref{fig5}(a), $T_\mathrm{c}$ of Hf$_{10}$Nb$_{25}$Sc$_{25}$Ti$_{20}$Zr$_{20}$ slightly increases from the as-cast state to the sample annealed at 400 \hspace{1mm}$^{\circ}$C and is markedly enhanced upon heat treatment at 500 \hspace{1mm}$^{\circ}$C. 
$T_\mathrm{c}$ reaches its maximum in the sample annealed at 600 \hspace{1mm}$^{\circ}$C, followed by a slight decrease after annealing at 800 \hspace{1mm}$^{\circ}$C (see Fig.\hspace{1 mm}\ref{fig5}(a) and Fig.\hspace{1 mm}\ref{fig6}(a)). 
The $C_\mathrm{p}$ data for each Hf$_{10}$Nb$_{25}$Sc$_{25}$Ti$_{20}$Zr$_{20}$ sample indicate the bulk nature of the superconducting transition (Fig.\hspace{1 mm}\ref{fig5}(b)). 
The agreement among the transition temperatures determined from $C_\mathrm{p}$, $M$, and $\rho$ measurements is relatively satisfactory for the samples annealed at 500 \hspace{1mm}$^{\circ}$C, 600 \hspace{1mm}$^{\circ}$C, and 800 \hspace{1mm}$^{\circ}$C. 
In contrast, for the as-cast and 400 \hspace{1mm}$^{\circ}$C-annealed samples, the onset temperature of the superconducting transition observed in $\rho$ or $M$ is 2–3 K higher than that in $C_\mathrm{p}$.
To investigate this discrepancy, the temperature dependence of the electronic specific heat $C_\mathrm{el}$($T$) was obtained by subtracting the phonon contribution represented by $\beta T^{3}$, and the resulting $C_\mathrm{el}$($T$)/$\gamma$$T$ ($\gamma$: electronic specific heat coefficient) was compared with $\rho$($T$) and $M$($T$), as shown in Figs.\hspace{1mm}S5(a) and (b) of the Supplementary Material. 
This comparison reveals that a weak anomaly in $C_\mathrm{el}$($T$)/$\gamma$$T$ emerges below the temperature at which $\rho$($T$) or $M$($T$) exhibits a sharp superconducting transition, suggesting an inhomogeneous superconducting state. 
This hypothesis is supported by SEM images of the as-cast and 400 \hspace{1mm}$^{\circ}$C annealed samples. 
In both, the microstructure is characterized by dendritic bcc and interdendritic hcp phases, with partial eutectic regions present along grain boundaries. 
As discussed later, the eutectic structure contributes to the enhancement of $T_\mathrm{c}$. 
Furthermore, the weave-like microstructure within the bcc phase may act as a precursor to eutectic formation, as thermal annealing promotes eutectic development within the bcc matrix. 
Therefore, the partial eutectic structure and weave-like microstructure likely contribute to the gradual enhancement of $T_\mathrm{c}$, manifested as the weak anomaly in $C_\mathrm{el}$($T$)/$\gamma$$T$ observed at 5–7 K.

Although the $T_\mathrm{c}$ of Hf$_{5}$Nb$_{45}$Sc$_{20}$Ti$_{15}$Zr$_{15}$ remains nearly unchanged between the as-cast state and the 400 \hspace{1mm}$^{\circ}$C-annealed sample, an increase in the annealing temperature from 400 \hspace{1mm}$^{\circ}$C to 600 \hspace{1mm}$^{\circ}$C induces a systematic enhancement of $T_\mathrm{c}$ (Fig.\hspace{1 mm}\ref{fig5}(c) and Fig.\hspace{1 mm}\ref{fig6}(b)). 
Further elevation of the annealing temperature up to 800 \hspace{1mm}$^{\circ}$C exerts only a minimal effect on $T_\mathrm{c}$. 
The annealing temperature dependence of $T_\mathrm{c}$ determined by $M$ or $\rho$ measurements is consistent with that derived from $C_\mathrm{p}$($T$), all indicating bulk superconducting transitions, as demonstrated in  Fig.\hspace{1 mm}\ref{fig5}(d).
For the Hf$_{5}$Nb$_{45}$Sc$_{10}$Ti$_{5}$Zr$_{35}$ system, the $T_\mathrm{c}$ of the as-cast sample decreases with annealing up to 400 \hspace{1mm}$^{\circ}$C before exhibiting a systematic increase with annealing above 400 \hspace{1mm}$^{\circ}$C (Fig.\hspace{1 mm}\ref{fig5}(e) and Fig.\hspace{1 mm}\ref{fig6}(c)). 
The $M$($T$) curve for the 500 \hspace{1mm}$^{\circ}$C annealed sample shows a broad, two-step diamagnetic transition, in contrast to the sharp resistivity drop observed in $\rho$($T$). 
This distinctive feature is prominently reflected in the $C_\mathrm{p}$($T$) data in Fig.\hspace{1 mm}\ref{fig5}(f). 
Unlike other samples exhibiting sharp bulk transitions, the 500 \hspace{1mm}$^{\circ}$C annealed sample displays an extremely broad transition. 
This sample lies near the boundary between phase-segregated bcc phases devoid of eutectic structure and those containing a minor eutectic component, implying the presence of phase instability. 
Consequently, the broad transition in $C_\mathrm{p}$($T$) is likely attributable to this phase instability.

\begin{table}
\caption{$T_\mathrm{c}$, $\mu_{0}H_\mathrm{c1}$(0), $\mu_{0}H_\mathrm{c2}$(0), $\xi_\mathrm{GL}$(0), $\lambda_\mathrm{GL}$(0), $\kappa_\mathrm{GL}$, $\alpha_\mathrm{M}$, $\lambda_\mathrm{SO}$, $\mu_{0}H_\mathrm{c2}^\mathrm{orb}$, and $\mu_{0}H_\mathrm{c2}^\mathrm{Pauli}$ of prepared samples. $T_\mathrm{c}$ is defined as the midpoint of specific heat jump, except for the as-cast and 400 \hspace{1mm}$^{\circ}$C annealed samples of Hf$_{10}$Nb$_{25}$Sc$_{25}$Ti$_{20}$Zr$_{20}$ and the 500 \hspace{1mm}$^{\circ}$C annealed sample of Hf$_{5}$Nb$_{45}$Sc$_{10}$Ti$_{5}$Zr$_{35}$. For these three samples, $T_\mathrm{c}$ is determined by the midpoint of the diamagnetic transition. $\mu_{0}H_\mathrm{c2}^\mathrm{Pauli}$ is calculated using the relation 1.84$T_\mathrm{c}^{\rho\hspace{0.5mm} 50\%}$, where $T_\mathrm{c}^{\rho\hspace{0.5mm} 50\%}$ is the superconducting critical temperature defined by the 50 \hspace{1mm}\% criterion of the resistivity drop in the zero-field $\rho$($T$). The sets of $T_\mathrm{c}^{\rho\hspace{0.5mm} 50\%}$ (K) values for the (as-cast, 400 \hspace{1mm}$^{\circ}$C, 500 \hspace{1mm}$^{\circ}$C, 600 \hspace{1mm}$^{\circ}$C, 800 \hspace{1mm}$^{\circ}$C) samples are (7.35, 7.56, 8.45, 9.01, 8.68) for Hf$_{10}$Nb$_{25}$Sc$_{25}$Ti$_{20}$Zr$_{20}$, (8.40, 8.37, 8.80, 9.51, 9.60) for Hf$_{5}$Nb$_{45}$Sc$_{20}$Ti$_{15}$Zr$_{15}$, and (8.65, 8.35, 9.34, 9.98, 10.1) for Hf$_{5}$Nb$_{45}$Sc$_{10}$Ti$_{5}$Zr$_{35}$, respectively.}\label{tab2}%
\scriptsize
\begin{tabular}{ccccccccccc}
\hline
sample  & $T_\mathrm{c}$ & $\mu_{0}H_\mathrm{c1}$(0) & $\mu_{0}H_\mathrm{c2}$(0) & $\xi_\mathrm{GL}$(0) & $\lambda_\mathrm{GL}$(0) &  $\kappa_\mathrm{GL}$ & $\alpha_\mathrm{M}$ & $\lambda_\mathrm{SO}$ & $\mu_{0}H_\mathrm{c2}^\mathrm{orb}$ & $\mu_{0}H_\mathrm{c2}^\mathrm{Pauli}$ \\
  & (K) & (mT) & (T) & (nm) & (nm) & & & & (T) & (T) \\
\hline
Hf$_{10}$Nb$_{25}$Sc$_{25}$Ti$_{20}$Zr$_{20}$ & & & & & & & & & & \\
as-cast & 6.83 & 23.6 & 12.8 & 5.1 & 154 & 30 & 1.49 & 4.0 & 14.2 & 13.5 \\ 
400 \hspace{1mm}$^{\circ}$C & 7.15 & 24.7 & 13.1 & 5.0 & 151 & 30 & 1.49 & 4.0 & 14.6 & 13.9 \\ 
500 \hspace{1mm}$^{\circ}$C & 8.36 & 25.5 & 11.5 & 5.4 & 146 & 27 & 1.76 & 0.5 & 19.3 & 15.5 \\ 
600 \hspace{1mm}$^{\circ}$C & 8.75 & 27.3 & 13.3 & 5.0 & 142 & 29 & 1.22 & 4.0 & 14.3 & 16.6 \\ 
800 \hspace{1mm}$^{\circ}$C & 8.37 & 23.6 & 13.3 & 5.0 & 155 & 31 & 1.27 & 4.0 & 14.4 & 16.0 \\ 
Hf$_{5}$Nb$_{45}$Sc$_{20}$Ti$_{15}$Zr$_{15}$ & & & & & & & & & &\\
as-cast & 7.69 & 18.5 & 11.4 & 5.4 & 176 & 33 & 1.16 & 2.0 & 12.8 & 15.5 \\ 
400 \hspace{1mm}$^{\circ}$C & 7.64 & 17.8 & 11.2 & 5.4 & 180 & 33 & 1.00 & 2.0 & 10.9 & 15.4 \\ 
500 \hspace{1mm}$^{\circ}$C & 8.17 & 33.1 & 11.1 & 5.4 & 125 & 23 & 1.28 & 0.5 & 14.7 & 16.2 \\ 
600 \hspace{1mm}$^{\circ}$C & 9.28 & 32.5 & 11.0 & 5.5 & 126 & 23 & 0.93 & 4.0 & 11.5 & 17.5 \\ 
800 \hspace{1mm}$^{\circ}$C & 9.39 & 26.5 & 10.8 & 5.5 & 142 & 26 & 0.9 & 4.0 & 11.3 & 17.7 \\ 
Hf$_{5}$Nb$_{45}$Sc$_{10}$Ti$_{5}$Zr$_{35}$ & & & & & & & & & & \\
as-cast & 7.73 & 8.2 & 12.1 & 5.2 & 283 & 54 & 1.10 & 4.0 & 12.4 & 15.9 \\ 
400 \hspace{1mm}$^{\circ}$C & 7.39 & 7.0 & 11.9 & 5.3 & 310 & 59 & 1.11 & 4.0 & 12.1 & 15.4 \\ 
500 \hspace{1mm}$^{\circ}$C & 8.25 & 36.5 & 11.7 & 5.3 & 118 & 22 & 0.80 & - & 9.7 & 17.2 \\ 
600 \hspace{1mm}$^{\circ}$C & 9.61 & 31.9 & 10.0 & 5.7 & 126 & 22 & 0.65 & - & 8.4 & 18.4 \\ 
800 \hspace{1mm}$^{\circ}$C & 9.93 & 26.6 & 10.8 & 5.5 & 142 & 26 & 0.68 & - & 9.0 & 18.6 \\ 
\hline
\end{tabular}
\end{table}

The lower critical field $H_\mathrm{c1}$ of each sample was evaluated from the isothermal $M$-$H$ curves at low magnetic fields, as presented in Figs.\hspace{1mm}S6-S8 of the Supplementary Material. 
The $H_\mathrm{c1}$ value is defined at the field where the magnetization deviates from the initial linear slope (indicated by a dashed line in each figure). 
The extracted $H_\mathrm{c1}$ values are plotted as a function of temperature in Figs.\hspace{1 mm}\ref{fig6}(d)-(f) and analyzed using the Ginzburg–Landau expression:
\begin{equation}
H_\mathrm{c1}(T)=H_\mathrm{c1}(0)\left(1-\left(\frac{T}{T_\mathrm{c}}\right)^{2}\right),
\label{eq:hc1}
\end{equation}
where $H_\mathrm{c1}$(0) denotes the lower critical field at zero temperature. 
All derived parameters are summarized in Table \ref{tab2}. 
While thermal annealing exhibits a minimal effect on $H_\mathrm{c1}$(0) in the Hf$_{10}$Nb$_{25}$Sc$_{25}$Ti$_{20}$Zr$_{20}$ system, heat treatment above 400 \hspace{1mm}$^{\circ}$C tends to enhance $H_\mathrm{c1}$(0) in the Hf$_{5}$Nb$_{45}$Sc$_{20}$Ti$_{15}$Zr$_{15}$ and Hf$_{5}$Nb$_{45}$Sc$_{10}$Ti$_{5}$Zr$_{35}$ systems.

Figures \ref{fig6}(g)-(i) summarize the temperature dependence of the upper critical field $H_\mathrm{c2}$ for all systems, derived from the 50 \hspace{1mm}\% resistivity drop criterion in $\rho$($T$) under external magnetic fields ranging from 0 to 9 T (see Figs.\hspace{1mm}S9-S11 in the Supplementary Material). 
The temperature dependence of $\mu_{0}H_\mathrm{c2}$ is analyzed using the Werthamer–Helfand–Hohenberg (WHH) model\cite{WHH:PR1966}:
\begin{equation}
\footnotesize
\mathrm{ln}\frac{1}{t}=(\frac{1}{2}+\frac{i\lambda_\mathrm{SO}}{4\gamma_\mathrm{WHH}})\psi\left(\frac{1}{2}+\frac{h+\lambda_\mathrm{SO}/2+i\gamma_\mathrm{WHH}}{2t} \right)+(\frac{1}{2}-\frac{i\lambda_\mathrm{SO}}{4\gamma_\mathrm{WHH}})\psi\left(\frac{1}{2}+\frac{h+\lambda_\mathrm{SO}/2-i\gamma_\mathrm{WHH}}{2t} \right)-\psi\left(\frac{1}{2}\right),
\label{eq:hc2}
\end{equation}
where $t=T/T_\mathrm{c}$, $\lambda_\mathrm{SO}$ is the spin-orbital scattering parameter, $\gamma_\mathrm{WHH}=[(\alpha_\mathrm{M}h)^{2}-(\lambda_\mathrm{SO}/2)^{2}]^{1/2}$ ($\alpha_\mathrm{M}$: Maki parameter), and $\psi(x)$ is the digamma function.
In this model, $h$ is defined as 
\begin{equation}
h=\frac{4H_\mathrm{c2}(T)}{\pi^{2}T_\mathrm{c}(-dH_\mathrm{c2}(T)/dT)_{T=T\mathrm{c}}}.
\label{eq:h-def}
\end{equation}
Once $\alpha_\mathrm{M}$ is determined, the orbital-limited upper critical field $\mu_{0}H_\mathrm{c2}^\mathrm{orb}$ is evaluated using the expression $\frac{\sqrt{2}\mu_{0}H_\mathrm{c2}^\mathrm{orb}}{\mu_{0}H_\mathrm{c2}^\mathrm{Pauli}}$, where $\mu_{0}H_\mathrm{c2}^\mathrm{Pauli}$ (=1.84$T_\mathrm{c}$) denotes the Pauli limiting field.
For this calculation, $T_\mathrm{c}^{\rho\hspace{0.5mm} 50\%}$ —the critical temperature determined from the 50 \hspace{1mm}\% resistivity drop in zero field— is used (see also the caption of Table \ref{tab2}). 
Table \ref{tab2} compiles all estimated parameters.
For the 500 \hspace{1mm}$^{\circ}$C, 600 \hspace{1mm}$^{\circ}$C, 800 \hspace{1mm}$^{\circ}$C samples of the Hf$_{5}$Nb$_{45}$Sc$_{10}$Ti$_{5}$Zr$_{35}$ system, the WHH model fails to accurately fit the $\mu_{0}H_\mathrm{c2}$($T$) data, as exemplified in Fig.\hspace{1mm}S12 of the Supplementary Material.
In these cases, the commonly adopted phenomenological formula $H_\mathrm{c2}(T)=H_\mathrm{c2}(0)\left(\frac{1-(T/T_\mathrm{c})^{2}}{1+(T/T_\mathrm{c})^{2}}\right)$ provides a better fit to the experimental data. 
The value of $\mu_{0}H_\mathrm{c2}^\mathrm{orb}$ is then calculated via $\mu_{0}H_\mathrm{c2}^\mathrm{orb}=-0.693T_\mathrm{c}\frac{dH_\mathrm{c2}(T)}{dT}|_{T=T_\mathrm{c}}$\cite{Jangid:APL2024}.
Comparison of the zero-temperature upper critical field $\mu_{0}H_\mathrm{c2}$(0) indicates that the Ti-rich Hf$_{10}$Nb$_{25}$Sc$_{25}$Ti$_{20}$Zr$_{20}$ system possesses a relatively high $\mu_{0}H_\mathrm{c2}$(0), consistent with previous observations in Ti-rich bcc HEAs\cite{Sobota:AM2025}. 
Samples of Hf$_{10}$Nb$_{25}$Sc$_{25}$Ti$_{20}$Zr$_{20}$ subjected to annealing below 500 \hspace{1mm}$^{\circ}$C exhibit $\alpha_\mathrm{M}>\sqrt{2}$, implying that $\mu_{0}H_\mathrm{c2}^\mathrm{orb}$ exceeds $\mu_{0}H_\mathrm{c2}^\mathrm{Pauli}$. 
A similar characteristic is reported in NbScTiZr, where lattice strain serves as a strong pinning center, suppressing Cooper pair breaking by the Lorentz force\cite{Kitagawa:MTC2024}. 
Given the comparable chemical composition, it is plausible that lattice strain also contributes to the enhancement of $\mu_{0}H_\mathrm{c2}^\mathrm{orb}$ in Hf$_{10}$Nb$_{25}$Sc$_{25}$Ti$_{20}$Zr$_{20}$. 
In contrast, most samples in the Hf$_{5}$Nb$_{45}$Sc$_{20}$Ti$_{15}$Zr$_{15}$ and Hf$_{5}$Nb$_{45}$Sc$_{10}$Ti$_{5}$Zr$_{35}$ systems exhibit relatively lower $\alpha_\mathrm{M}$ values, indicating that $\mu_{0}H_\mathrm{c2}$(0) is governed by $\mu_{0}H_\mathrm{c2}^\mathrm{orb}$.

The Ginzburg–Landau coherence length $\xi_\mathrm{GL}$(0) was calculated using the relation $\xi_\mathrm{GL}(0)=\sqrt{\frac{\Phi_{0}}{2\pi\mu_{0}H_\mathrm{c2}(0)}}$, where $\Phi_{0}=2.07\times 10^{-15}$ Wb is the magnetic flux quantum. 
The corresponding $\xi_\mathrm{GL}$(0) values are listed in Table \ref{tab2}. 
The range of 5.0 to 5.7 nm is comparable to typical HEAs (e.g., NbReHfZrTi: 6.1 nm\cite{Marik:JALCOM2018}, Ta$_{1/6}$Nb$_{2/6}$Hf$_{1/6}$Zr$_{1/6}$Ti$_{1/6}$: 5.2 nm\cite{Kim:AM2020}). 
The magnetic penetration depth $\lambda_\mathrm{GL}$(0), determined using $\mu_{0}H_\mathrm{c1}(0)=\frac{\Phi_{0}}{4\pi\lambda_\mathrm{GL}(0)^{2}}\mathrm{ln}\frac{\lambda_\mathrm{GL}(0)}{\xi_\mathrm{GL}(0)}$, is also included in Table \ref{tab2}. 
Consequently, the Ginzburg–Landau parameter $\kappa_\mathrm{GL}$=$\lambda_{GL}(0)/\xi_{GL}(0)$ was calculated and tabulated for each sample. 
All $\kappa_\mathrm{GL}$ values exceeding 1/$\sqrt{2}$ confirm that the investigated alloy systems are classified as type-II superconductors.

Here, we discuss the relationship between the $C_\mathrm{p}$($T$) behavior and the superconducting parameters $T_\mathrm{c}$ and $\mu_{0}H_\mathrm{c2}$(0).
Figures \ref{fig5}(b), (d), and (f) indicate that the specific-heat jump at $T_\mathrm{c}$ becomes sharper with increasing $T_\mathrm{c}$ for a fixed chemical composition, except in the case of Hf$_{5}$Nb$_{45}$Sc$_{10}$Ti$_{5}$Zr$_{35}$. 
For Hf$_{5}$Nb$_{45}$Sc$_{10}$Ti$_{5}$Zr$_{35}$, both the as-cast and 400 \hspace{1mm}$^{\circ}$C-annealed samples consist solely of a bcc structure, and the degree of inhomogeneity is relatively small, which may account for the sharper specific-heat jump even in the lower-$T_\mathrm{c}$ sample. 
In Hf$_{10}$Nb$_{25}$Sc$_{25}$Ti$_{20}$Zr$_{20}$ and Hf$_{5}$Nb$_{45}$Sc$_{20}$Ti$_{15}$Zr$_{15}$, the low-temperature-annealed samples with reduced $T_\mathrm{c}$ likely exhibit significant inhomogeneity arising from the fine coexistence of bcc and hcp microstructures formed during nonequilibrium solidification. 
High-temperature annealing generally induces microstructural coarsening and mitigates inhomogeneity, thereby giving rise to a sharper superconducting transition. 
Thus, the correlation between $C_\mathrm{p}$($T$) behavior and $T_\mathrm{c}$ is intrinsically linked to the issue of sample homogeneity. 
Regarding the relationship between $C_\mathrm{p}$($T$) behavior and $\mu_{0}H_\mathrm{c2}$(0), the magnitude of $\mu_{0}H_\mathrm{c2}$(0) appears independent of sample homogeneity, as reflected in the sharpness of the specific-heat jump (see also Table \ref{tab2}). 
For instance, although Hf$_{5}$Nb$_{45}$Sc$_{10}$Ti$_{5}$Zr$_{35}$ annealed at 500 \hspace{1mm}$^{\circ}$C exhibits a considerably broadened $C_\mathrm{p}$($T$) transition, $\mu_{0}H_\mathrm{c2}$(0) remains unaffected. 
We evaluated $\mu_{0}H_\mathrm{c2}$(0) from $\rho$($T$), which likely probes regions exhibiting the highest $\mu_{0}H_\mathrm{c2}$ even in inhomogeneous samples. 
The difference in measurement methodology may therefore account for the insensitivity of $\mu_{0}H_\mathrm{c2}$(0) to $C_\mathrm{p}$($T$) behavior.

Several bcc HEA superconductors fall within the strong-coupling regime, as evidenced by $\frac{2\Delta(0)}{k_\mathrm{B}T_\mathrm{c}}$ and $\frac{\Delta C_\mathrm{el}}{\gamma T_\mathrm{c}}$ values exceeding the BCS weak-coupling limits of 3.52 and 1.43, respectively\cite{Kozelj:PRL2014,Marik:JALCOM2018,Kim:AM2020,Rohr:PNAS2016,Sarkar:IM2022,Zeng:AQT2023}. 
Here, $\Delta(0)$ denotes the superconducting gap at zero temperature, $k_\mathrm{B}$ is the Boltzmann constant, $\Delta C_\mathrm{el}$ represents the specific heat jump at $T_\mathrm{c}$, and $\gamma$ is the electronic specific heat coefficient. 
The electronic specific heat $C_\mathrm{el}$ is obtained by subtracting the phonon contribution, expressed as $\beta T^{3}$, from $C_\mathrm{p}$. 
The $\gamma$ and $\beta$ values are determined by fitting $C_\mathrm{p}$($T$) in the normal state using the relation $\gamma T+\beta T^{3}$ (refer to Figs.\hspace{1mm}\ref{fig5}(b), (d), and (f)). 
The ($\gamma$ (mJ/mol$\cdot$K$^{2}$), $\beta$ (mJ/mol$\cdot$K$^{4}$)) values for the as-cast, 400 \hspace{1mm}$^{\circ}$C, 500 \hspace{1mm}$^{\circ}$C, 600 \hspace{1mm}$^{\circ}$C, and 800 \hspace{1mm}$^{\circ}$C samples are estimated to be (6.72, 0.153), (6.70, 0.153), (5.40, 0.133), (4.83, 0.132), and (5.10, 0.135) in Hf$_{10}$Nb$_{25}$Sc$_{25}$Ti$_{20}$Zr$_{20}$; (6.72, 0.153), (6.70, 0.153), (6.68, 0.132), (6.70, 0.130), and (6.65, 0.128) in Hf$_{5}$Nb$_{45}$Sc$_{20}$Ti$_{15}$Zr$_{15}$; and (7.62, 0.220), (6.51, 0.180), (6.18, 0.142), (5.80, 0.120), and (5.10, 0.132) in Hf$_{5}$Nb$_{45}$Sc$_{10}$Ti$_{5}$Zr$_{35}$, respectively.
For the as-cast and 400 \hspace{1mm}$^{\circ}$C annealed samples of Hf$_{5}$Nb$_{45}$Sc$_{10}$Ti$_{5}$Zr$_{35}$, which consist solely of the bcc phase, the Debye temperature $\theta_\mathrm{D}$ is estimated from the $\beta$ value using the relation $\theta_\mathrm{D}=\left(\frac{12\pi^{4}RN}{5\beta}\right)^{1/3}$, where $R$ is the gas constant and $N$ = 1 is the number of atoms per formula unit. 
The $\theta_\mathrm{D}$ values are 207 K and 221 K for the as-cast and 400 \hspace{1mm}$^{\circ}$C annealed samples, respectively.
To evaluate the $\frac{2\Delta(0)}{k_\mathrm{B}T_\mathrm{c}}$ ratio, $C_\mathrm{el}$ is calculated within the $\alpha$-model\cite{Padamsee:LTP1973,Johnston:SST2013} using the equation:
\begin{equation}
\frac{C_\mathrm{el}(t)}{\gamma T_\mathrm{c}}=\frac{6\alpha^{3}}{\pi^{2}t}\int_{0}^{\infty}f(1-f)\left(\frac{\tilde{E}^{2}}{t}-\frac{1}{2}\frac{d\tilde{\Delta}^{2}}{dt}\right)d\tilde{\epsilon},
\label{eq:alpha}
\end{equation}
where $\alpha$=$\Delta(0)/k_\mathrm{B}T_\mathrm{c}$ is an adjustable parameter, $\tilde{\epsilon}$=$\epsilon/\Delta(0)$ is the normalized energy, $\tilde{E}$=$\sqrt{\tilde{\epsilon}^{2}+\tilde{\Delta}^{2}}$ is the normalized energy of superconducting quasi-particle excitations, $f$ is the Fermi distribution function, and $\tilde{\Delta}$=$\Delta/\Delta(0)$ is the superconducting gap normalized to its zero-temperature value. 
The BCS weak-coupling limit corresponds to $\alpha$ = 1.76.

In the current analysis, all samples - except the as-cast and 400 \hspace{1mm}$^{\circ}$C annealed samples of Hf$_{5}$Nb$_{45}$Sc$_{10}$Ti$_{5}$Zr$_{35}$ - require an electronic specific heat contribution attributable to the normal state, even within the superconducting regime. 
Given that the as-cast and 400 \hspace{1mm}$^{\circ}$C annealed samples of Hf$_{5}$Nb$_{45}$Sc$_{10}$Ti$_{5}$Zr$_{35}$ exhibit solely the bcc phase, it is inferred that the hcp phase in the other samples remains in the normal state down to 2 K.
Accordingly, the electronic specific heat contribution from the hcp phase is modeled as $\gamma_\mathrm{hcp} T$, where $\gamma_\mathrm{hcp}$ denotes the hcp contribution to the total $\gamma$. 
Therefore, the temperature-dependent electronic specific heat $C_\mathrm{el}$($T$) is described by:
\begin{equation}
C_\mathrm{el}(t)=\frac{\gamma_\mathrm{bcc}}{\gamma}C_\mathrm{el\hspace{1mm}\alpha-model}(t)+\frac{\gamma_\mathrm{hcp}}{\gamma}T,
\label{eq:alpha2}
\end{equation}
where $\gamma_\mathrm{bcc}$ and $C_\mathrm{el\hspace{1mm}\alpha-model}(t)$ denote the bcc-phase contribution to $\gamma$ and the electronic specific heat calculated via the $\alpha$-model, respectively. 
$T_\mathrm{c}$ is defined as the midpoint of the specific heat jump. 
The discontinuities in $C_\mathrm{el}$ at $T_\mathrm{c}$ are determined by extrapolating data below and above $T_\mathrm{c}$, as shown by dashed lines in Figs.\hspace{1mm}S13-S15 of the Supplementary Material. 
In computing $\frac{\Delta C_\mathrm{el}}{\gamma T_\mathrm{c}}$, $\gamma$ is replaced by $\gamma_\mathrm{bcc}$. 
It is noted that $\frac{2\Delta(0)}{k_\mathrm{B}T_\mathrm{c}}$ and $\frac{\Delta C_\mathrm{el}}{\gamma T_\mathrm{c}}$ are not evaluated for the as-cast and 400 \hspace{1mm}$^{\circ}$C annealed samples of Hf$_{10}$Nb$_{25}$Sc$_{25}$Ti$_{20}$Zr$_{20}$ and the 500 \hspace{1mm}$^{\circ}$C annealed sample of Hf$_{5}$Nb$_{45}$Sc$_{10}$Ti$_{5}$Zr$_{35}$, due to their extremely broad transitions, which hinder accurate parameter extraction.

\begin{figure}
\begin{center}
\includegraphics[width=1.1\linewidth]{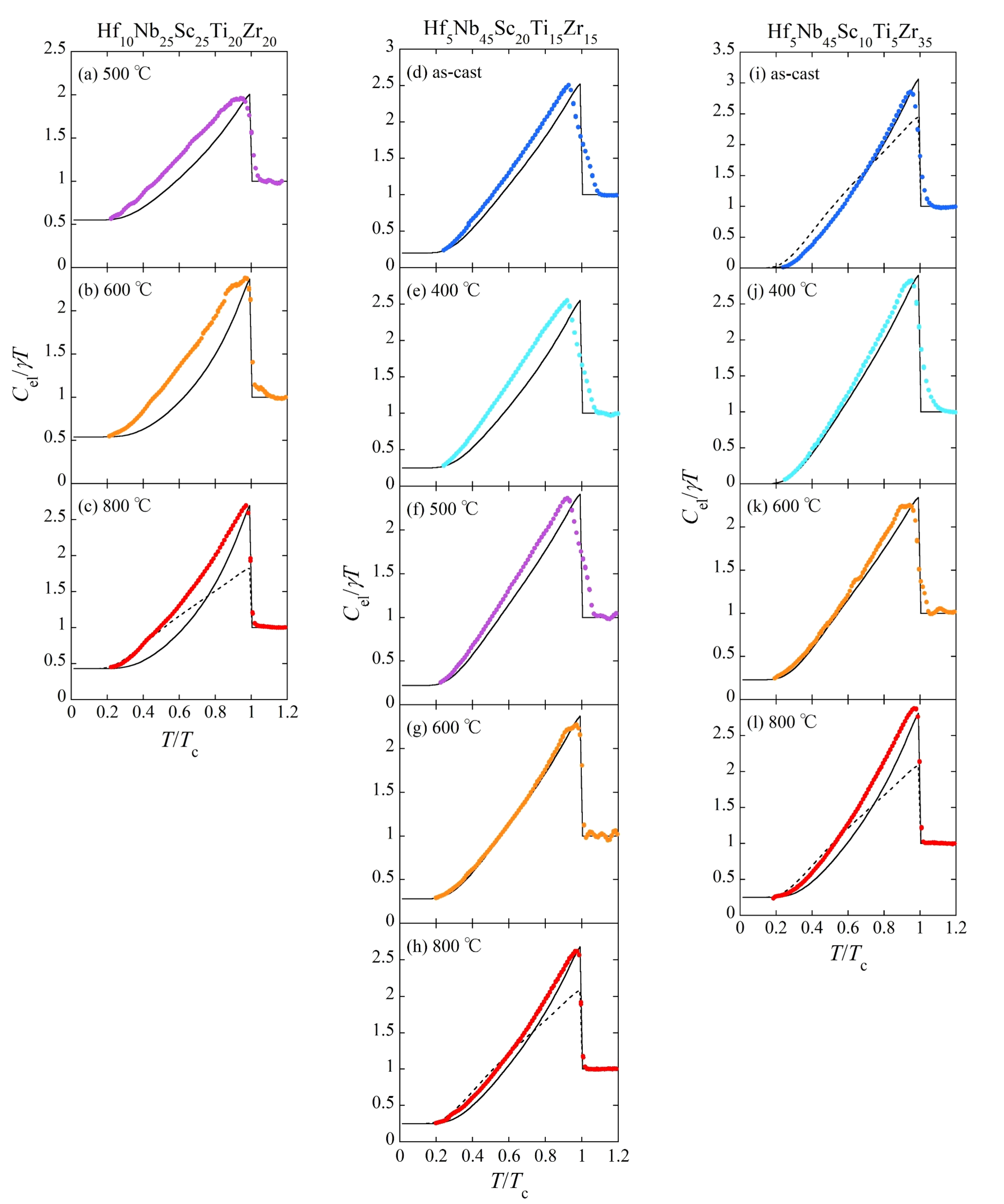}
\caption{\label{fig7}$C_\mathrm{el}/\gamma T$ vs. $T/T_\mathrm{c}$ plots for (a)-(c) Hf$_{10}$Nb$_{25}$Sc$_{25}$Ti$_{20}$Zr$_{20}$, (d)-(h) Hf$_{5}$Nb$_{45}$Sc$_{20}$Ti$_{15}$Zr$_{15}$, and (i)-(l) Hf$_{5}$Nb$_{45}$Sc$_{10}$Ti$_{5}$Zr$_{35}$ systems, respectively. The solid lines represent fitting results using Eq.\hspace{1mm}(\ref{eq:alpha2}), while the dashed lines correspond to fittings using Eq.\hspace{1mm}(\ref{eq:alpha2}) with $\alpha$=1.76.}
\end{center}
\end{figure}

Figures \ref{fig7}(a)-(l) depict the $C_\mathrm{el}/\gamma T$ vs. $T/T_\mathrm{c}$ plots for all studied systems. 
The solid lines represent fits obtained using Eq.\hspace{1mm}(\ref{eq:alpha2}), and the extracted parameters are summarized in Table \ref{tab3}. 
The resulting $\frac{2\Delta(0)}{k_\mathrm{B}T_\mathrm{c}}$ and $\frac{\Delta C_\mathrm{el}}{\gamma_\mathrm{bcc} T_\mathrm{c}}$ values, which both exceed the BCS weak-coupling limits of 3.52 and 1.43, underscore that the Hf–Nb–Sc–Ti–Zr system resides in the strong-coupling regime. 
As seen in Figs.\hspace{1mm}\ref{fig7}(a)-(c) for the Hf$_{10}$Nb$_{25}$Sc$_{25}$Ti$_{20}$Zr$_{20}$ samples, $\alpha$ values greater than 1.76 are necessary to reproduce the pronounced specific heat jumps. 
The weak-coupling value of $\alpha$ = 1.76 underestimates the jump magnitude, as illustrated by the dashed line in Fig.\hspace{1mm}\ref{fig7}(c).
Nonetheless, the overall fitting quality is limited by upper deviations of the experimental data from theoretical predictions, suggesting the presence of additional quasi-particle excitations in the superconducting state.
Similarly, the Hf$_{5}$Nb$_{45}$Sc$_{20}$Ti$_{15}$Zr$_{15}$ system also requires enhanced $\alpha$ values to reproduce $C_\mathrm{el}$ ($T$) (Figs.\hspace{1mm}\ref{fig7}(d)-(h)). 
For samples annealed below 500 \hspace{1mm}$^{\circ}$C, relatively large deviations between experimental and theoretical curves persist below and above $T_\mathrm{c}$, likely due to sample inhomogeneity stemming from insufficient annealing.
The as-cast and 400 \hspace{1mm}$^{\circ}$C annealed samples of Hf$_{5}$Nb$_{45}$Sc$_{10}$Ti$_{5}$Zr$_{35}$, which lack an hcp phase, require no $\gamma_\mathrm{hcp}$ contribution. 
For these samples, the strong-coupling $\alpha$-model effectively reproduces the experimental data (Figs.\hspace{1mm}\ref{fig7}(i) and (j)). 
The weak-coupling BCS value of $\alpha$ = 1.76 is inadequate to describe the temperature dependence. 
In the 800 \hspace{1mm}$^{\circ}$C annealed sample, a value of $\alpha$ exceeding 1.76 is again necessary. 
However, minor upper deviations in the experimental data, akin to those in the Hf$_{10}$Nb$_{25}$Sc$_{25}$Ti$_{20}$Zr$_{20}$ samples, are observed (Fig.\hspace{1mm}\ref{fig7}(l)).

\begin{table}
\caption{$\alpha$, $\frac{\gamma_\mathrm{bcc}}{\gamma}$, $\frac{\gamma_\mathrm{hcp}}{\gamma}$, $\frac{2\Delta(0)}{k_\mathrm{B}T_\mathrm{c}}$, $\frac{\Delta C_\mathrm{el}}{\gamma_\mathrm{bcc} T_\mathrm{c}}$, $\omega_\mathrm{ln}$, and $\lambda_\mathrm{ep}$ of Hf-Nb-Sc-Ti-Zr systems.}\label{tab3}%
\begin{tabular}{cccccccc}
\hline
sample  & $\alpha$ & $\frac{\gamma_\mathrm{bcc}}{\gamma}$ & $\frac{\gamma_\mathrm{hcp}}{\gamma}$ & $\frac{2\Delta(0)}{k_\mathrm{B}T_\mathrm{c}}$ & $\frac{\Delta C_\mathrm{el}}{\gamma_\mathrm{bcc} T_\mathrm{c}}$ & $\omega_\mathrm{ln}$ (K) & $\lambda_\mathrm{ep}$\\
\hline
Hf$_{10}$Nb$_{25}$Sc$_{25}$Ti$_{20}$Zr$_{20}$ & & & & & & & \\
500 \hspace{1mm}$^{\circ}$C & 2.2 & 0.45 & 0.55 & 4.4 & 2.19 & 83.8 & 1.44 \\
600 \hspace{1mm}$^{\circ}$C & 2.55 & 0.46 & 0.54 & 5.1 & 3.07 & 44.8 & 4.19 \\ 
800 \hspace{1mm}$^{\circ}$C & 2.55 & 0.57 & 0.43 & 5.1 & 3.13 & 42.6 & 4.26 \\ 
Hf$_{5}$Nb$_{45}$Sc$_{20}$Ti$_{15}$Zr$_{15}$ & & & & & & &\\
as-cast & 2.02 & 0.8 & 0.2 & 4.04 & 2.18 & 92.5 & 1.24 \\ 
400 \hspace{1mm}$^{\circ}$C & 2.11 & 0.75 & 0.25 & 4.22 & 2.12 & 87.2 & 1.29 \\ 
500 \hspace{1mm}$^{\circ}$C & 1.97 & 0.78 & 0.22 & 3.94 & 1.91 & 128 & 1.04 \\ 
600 \hspace{1mm}$^{\circ}$C & 2.02 & 0.72 & 0.28 & 4.04 & 1.8 & 149 & 1.03 \\ 
800 \hspace{1mm}$^{\circ}$C & 2.2 & 0.75 & 0.25 & 4.4 & 2.33 & 87.7 & 1.54 \\ 
Hf$_{5}$Nb$_{45}$Sc$_{10}$Ti$_{5}$Zr$_{35}$ & & & & & & & \\
as-cast & 2.11 & 1.0 & 0.0 & 4.22 & 2.04 & 92.8 & 1.24 \\ 
400 \hspace{1mm}$^{\circ}$C & 2.02 & 1.0 & 0.0 & 4.04 & 1.75 & 124 & 1.00 \\ 
600 \hspace{1mm}$^{\circ}$C & 1.94 & 0.77 & 0.23 & 3.88 & 1.68 & 203 & 0.88 \\ 
800 \hspace{1mm}$^{\circ}$C & 2.29 & 0.75 & 0.25 & 4.58 & 2.68 & 73.6 & 2.01 \\ 
\hline
\end{tabular}
\end{table}

\subsection{Critical current density}
\begin{figure}
\begin{center}
\includegraphics[width=1\linewidth]{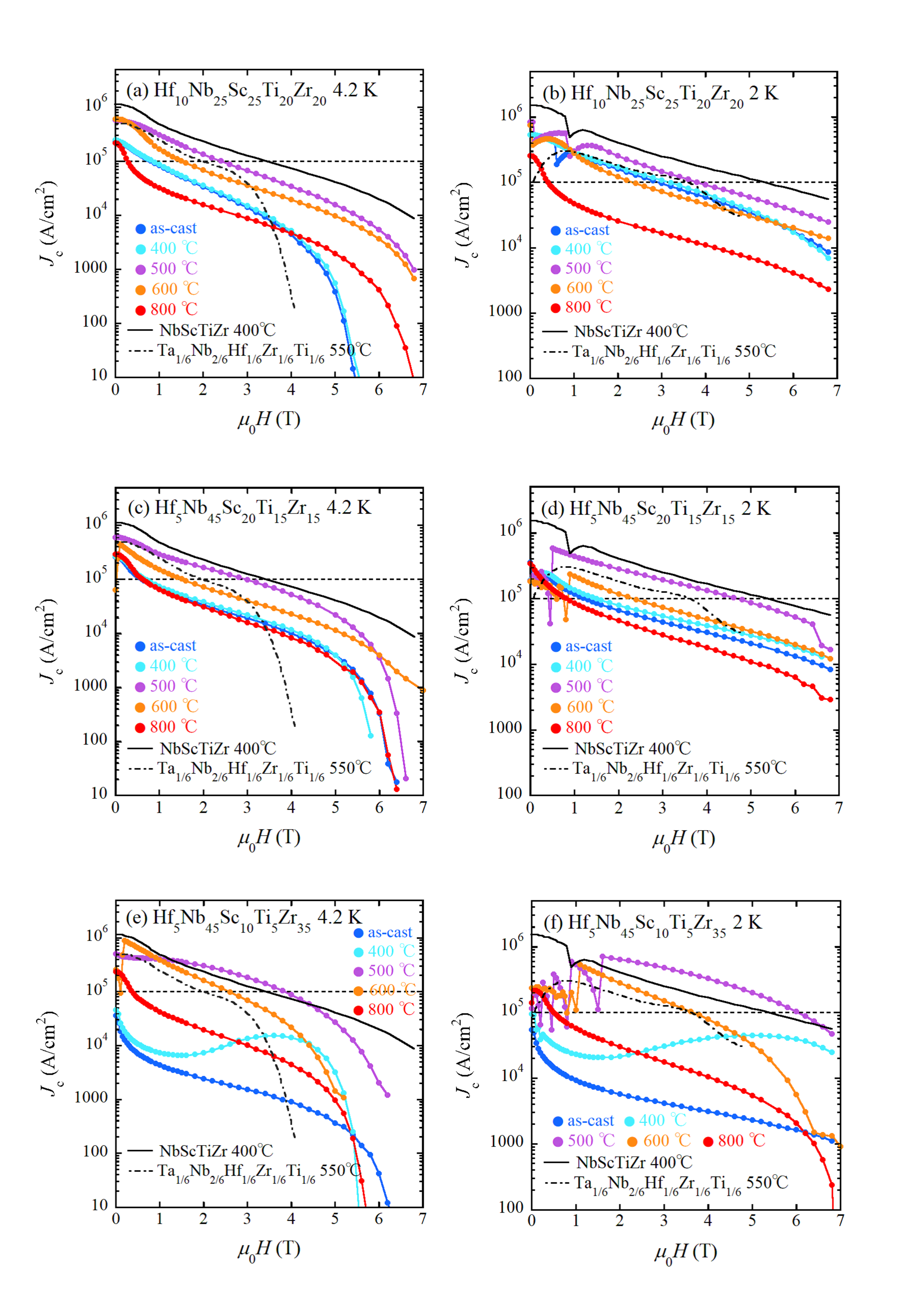}
\caption{\label{fig8} (a) and (b) Magnetic field dependence of $J_\mathrm{c}$ for Hf$_{10}$Nb$_{25}$Sc$_{25}$Ti$_{20}$Zr$_{20}$ at 4.2 and 2 K, respectively. (c) and (d) Magnetic field dependence of $J_\mathrm{c}$ for Hf$_{5}$Nb$_{45}$Sc$_{20}$Ti$_{15}$Zr$_{15}$ at 4.2 and 2 K, respectively. (e) and (f) Magnetic field dependence of $J_\mathrm{c}$ for Hf$_{5}$Nb$_{45}$Sc$_{10}$Ti$_{5}$Zr$_{35}$ at 4.2 and 2 K, respectively. The dashed lines indicate the practical benchmark level for superconducting wire applications. The solid and dash-dotted lines represent $J_\mathrm{c}$ values of NbScTiZr annealed at 400 \hspace{1mm}$^{\circ}$C\cite{Kitagawa:EPJB2025} and Ta$_{1/6}$Nb$_{2/6}$Hf$_{1/6}$Zr$_{1/6}$Ti$_{1/6}$ annealed at 550 \hspace{1mm}$^{\circ}$C\cite{Kim:JMST2024}, respectively.}
\end{center}
\end{figure}

$J_\mathrm{c}$ is calculated using the critical state model, expressed as follows\cite{Peterson:JAP1990}:
\begin{equation}
J_\mathrm{c}=\frac{3\Delta M}{\frac{3}{2}a\left(1-\frac{a}{3b}\right)},
\label{eq:Jc}
\end{equation}
where $\Delta M$ is the width of the isothermal $M$–$H$ hysteresis loop at an applied magnetic field $H$, and $a$ and $b$ ($b>a$) denote the sample dimensions perpendicular to $H$. 
The experimental $M$–$H$ hysteresis loops at 4.2 and 2 K for all samples are presented in Figs.\hspace{1mm}S16(a)-(f) in the Supplementary Material. 
Some figures exhibit abrupt changes in magnetization $M$, more pronounced at the lower temperature of 2 K. 
This behavior, commonly observed in various superconductors, including other bcc HEAs\cite{Jung:NC2022,Gao:APL2022,Dou:PhysicaC2001}, is attributed to flux jumps, indicative of strong flux pinning. 
The field-dependent $J_\mathrm{c}$ values, determined via the critical state model, are illustrated in Figs.\hspace{1mm}\ref{fig8}(a)-(f). 
Dashed lines denote the practical threshold of 10$^{5}$ A/cm$^{2}$, typically considered the benchmark for superconducting wire applications\cite{Jung:NC2022,Larbalestier:Nature2001,Komori:APL2002}. 
For comparison, $J_\mathrm{c}$ data of NbScTiZr annealed at 400 \hspace{1mm}$^{\circ}$C (solid lines)\cite{Kitagawa:EPJB2025} and Ta$_{1/6}$Nb$_{2/6}$Hf$_{1/6}$Zr$_{1/6}$Ti$_{1/6}$ annealed at 550 \hspace{1mm}$^{\circ}$C (dash-dotted lines)\cite{Kim:JMST2024} are also shown.

In both Hf$_{10}$Nb$_{25}$Sc$_{25}$Ti$_{20}$Zr$_{20}$ and Hf$_{5}$Nb$_{45}$Sc$_{20}$Ti$_{15}$Zr$_{15}$ systems, $J_\mathrm{c}$ values at 4.2 K and 2 K increase with annealing temperature up to 500 \hspace{1mm}$^{\circ}$C across the entire magnetic field range. 
At each temperature, the samples annealed at 500 \hspace{1mm}$^{\circ}$C exhibit superior performance, with $J_\mathrm{c}$ surpassing 10$^{5}$ A/cm$^{2}$ up to approximately 2.5 T (at 4.2 K) and 4 T (at 2 K) for Hf$_{10}$Nb$_{25}$Sc$_{25}$Ti$_{20}$Zr$_{20}$, and 3 T (at 4.2 K) and 5 T (at 2 K) for Hf$_{5}$Nb$_{45}$Sc$_{20}$Ti$_{15}$Zr$_{15}$. 
Heat treatments above 500 \hspace{1mm}$^{\circ}$C lead to a systematic degradation of $J_\mathrm{c}$ across the examined field range. 
The annealing temperature dependence of $J_\mathrm{c}$ correlates well with the variation in lattice parameter, as shown in Fig.\hspace{1mm}\ref{fig1}(d), which reaches a minimum at $\sim$ 500 \hspace{1mm}$^{\circ}$C, signifying enhanced lattice strain. 
As reported for Ta$_{1/6}$Nb$_{2/6}$Hf$_{1/6}$Zr$_{1/6}$Ti$_{1/6}$ and (TaNb)$_{0.7}$(HfZrTi)$_{0.5}$, increased lattice strain can enhance $J_\mathrm{c}$ via strengthened flux pinning\cite{Kitagawa:EPJB2025,Kim:JMST2024,Gao:APL2022}. 
Notably, the 500 \hspace{1mm}$^{\circ}$C annealed Hf$_{10}$Nb$_{25}$Sc$_{25}$Ti$_{20}$Zr$_{20}$ sample partially contains a fine eutectic structure with $\sim$ 80 nm spacing. 
As previously observed in NbScTiZr, such fine eutectic structures serve as effective flux pinning centers\cite{Seki:JSNM2023,Kitagawa:MTC2024}.
However, in Hf$_{10}$Nb$_{25}$Sc$_{25}$Ti$_{20}$Zr$_{20}$, this structure does not uniformly permeate the sample, likely limiting the flux pinning effect, reflected in the relatively lower $J_\mathrm{c}$ compared to NbScTiZr, as seen in  Figs.\hspace{1mm}\ref{fig8}(a) and (b). 
The $J_\mathrm{c}$ values of the 500 \hspace{1mm}$^{\circ}$C samples in both Hf$_{10}$Nb$_{25}$Sc$_{25}$Ti$_{20}$Zr$_{20}$ and Hf$_{5}$Nb$_{45}$Sc$_{20}$Ti$_{15}$Zr$_{15}$ are comparable to those reported for Ta$_{1/6}$Nb$_{2/6}$Hf$_{1/6}$Zr$_{1/6}$Ti$_{1/6}$ annealed at 550 \hspace{1mm}$^{\circ}$C\cite{Kim:JMST2024}, reinforcing the pivotal role of lattice strain in flux pinning.

The as-cast Hf$_{5}$Nb$_{45}$Sc$_{10}$Ti$_{5}$Zr$_{35}$ sample, lacking eutectic structure and exhibiting minimal lattice strain as suggested by Fig.\hspace{1mm}\ref{fig1}(d), displays markedly lower $J_\mathrm{c}$ (Figs.\hspace{1mm}\ref{fig8}(e) and (f)) in comparison to the other two systems. 
Although the 400 \hspace{1mm}$^{\circ}$C annealed sample also exhibits reduced $J_\mathrm{c}$, a fishtail effect is observed at higher magnetic fields. 
The enhancement in $J_\mathrm{c}$ at elevated fields likely underpins the significantly improved $J_\mathrm{c}$ across the entire magnetic field range for the 500 \hspace{1mm}$^{\circ}$C annealed sample, which lies near a phase instability, as discussed in the $C_\mathrm{p}$($T$) results. 
Remarkably, the $J_\mathrm{c}$ of the 500 \hspace{1mm}$^{\circ}$C annealed Hf$_{5}$Nb$_{45}$Sc$_{10}$Ti$_{5}$Zr$_{35}$ sample exceeds 10$^{5}$ A/cm$^{2}$ up to nearly 4 T (at 4.2 K) and 6 T (at 2 K), classifying it among the highest-performing HEAs in terms of $J_\mathrm{c}$. 
Similar to the other systems, further annealing at higher temperatures diminishes $J_\mathrm{c}$, and the annealing temperature dependence of the lattice parameter suggests that increased lattice strain contributes to the enhanced flux pinning observed at 500 \hspace{1mm}$^{\circ}$C. 
Although SEM analysis of this sample reveals a partially developed fine eutectic structure with $\sim$ 40 nm spacing, its contribution to flux pinning may be limited, akin to the case of Hf$_{10}$Nb$_{25}$Sc$_{25}$Ti$_{20}$Zr$_{20}$. 
This may explain the relatively lower $J_\mathrm{c}$ at low magnetic fields in the 500 \hspace{1mm}$^{\circ}$C annealed Hf$_{5}$Nb$_{45}$Sc$_{10}$Ti$_{5}$Zr$_{35}$ sample compared to NbScTiZr (Figs.\hspace{1mm}\ref{fig8}(e) and (f)). 
Nevertheless, at higher fields (1 T$<\mu_{0}H<$4.5 T at 4.2 K and $\mu_{0}H>$1.5 T at 2 K), the $J_\mathrm{c}$ of the 500 \hspace{1mm}$^{\circ}$C annealed sample surpasses those of both Ta$_{1/6}$Nb$_{2/6}$Hf$_{1/6}$Zr$_{1/6}$Ti$_{1/6}$ and NbScTiZr, suggesting an additional flux pinning mechanism beyond lattice strain. 
As previously noted, the specific heat displays an exceptionally broad anomaly following 500 \hspace{1mm}$^{\circ}$C annealing in Hf$_{5}$Nb$_{45}$Sc$_{10}$Ti$_{5}$Zr$_{35}$, implying a phase instability that could plausibly serve as an additional flux pinning source.

\begin{figure}
\begin{center}
\includegraphics[width=1.0\linewidth]{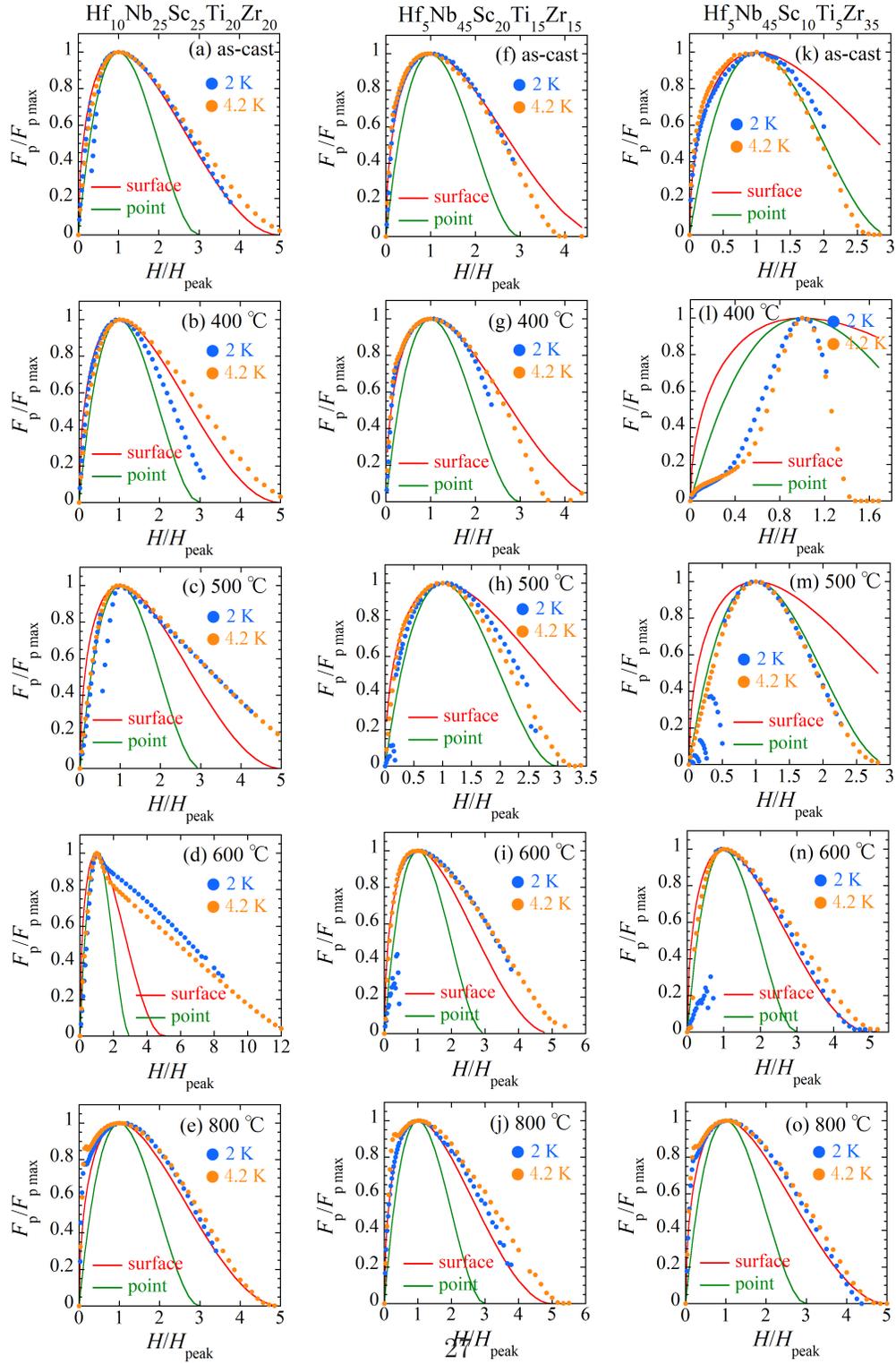}
\caption{\label{fig9} Normalized flux pinning force densities as a function of $H/H_\mathrm{peak}$ for (a)-(e) Hf$_{10}$Nb$_{25}$Sc$_{25}$Ti$_{20}$Zr$_{20}$, (f)-(j) Hf$_{5}$Nb$_{45}$Sc$_{20}$Ti$_{15}$Zr$_{15}$, and (k)-(o)  Hf$_{5}$Nb$_{45}$Sc$_{10}$Ti$_{5}$Zr$_{35}$, respectively.}
\end{center}
\end{figure}

To elucidate the flux pinning mechanism, the flux pinning force density ($F_\mathrm{p}$) was calculated using the relation $F_\mathrm{p}=\left| \mu_{0}\vec{H}\times\vec{J_\mathrm{c}}\right|$.
Subsequently, the normalized flux pinning force density, $f_\mathrm{p}$ ($=F_\mathrm{p}/F_\mathrm{p\hspace{0.5mm}max}$), was evaluated as a function of $H/H_\mathrm{peak}$, where $F_\mathrm{p\hspace{0.5mm}max}$ and $H_\mathrm{peak}$ represent the maximum value of $F_\mathrm{p}$ and the corresponding magnetic field, respectively. 
The experimental data were compared with representative theoretical models: 
\begin{equation}
f_\mathrm{p}=\frac{25}{16}\left(\frac{H}{H_\mathrm{peak}}\right)^{0.5}\left(1-\frac{1}{5}\frac{H}{H_\mathrm{peak}}\right)^{2}
\label{eq:nsp}
\end{equation}
for normal surface pinning and
\begin{equation}
f_\mathrm{p}=\frac{9}{4}\left(\frac{H}{H_\mathrm{peak}}\right)\left(1-\frac{1}{3}\frac{H}{H_\mathrm{peak}}\right)^{2}
\label{eq:npp}
\end{equation}
for normal point pinning\cite{Higuchi:PRB1999,Shigeta:PhysC2003,Cai:APL2013}. 
Figures \ref{fig9}(a)-(o) present $f_\mathrm{p}$ as a function of $H/H_\mathrm{peak}$ for all systems at 2 and 4.2 K, overlaid with theoretical curves corresponding to the surface and point pinning mechanisms.

In the Hf$_{10}$Nb$_{25}$Sc$_{25}$Ti$_{20}$Zr$_{20}$ system, the as-cast sample follows the surface pinning model, suggesting that grain boundaries between the bcc and hcp phases serve as primary flux pinning sites (Fig.\hspace{1mm}\ref{fig9}(a)). 
The samples annealed at 500 \hspace{1mm}$^{\circ}$C and 600 \hspace{1mm}$^{\circ}$C exhibit significant deviations from both models at higher magnetic fields, although their low-field behavior aligns with the point pinning mechanism (Figs.\hspace{1mm}\ref{fig9}(c) and (d)). 
Notably, the 600 \hspace{1mm}$^{\circ}$C annealed sample displays a shoulder-like feature centered around $H/H_\mathrm{peak}$ = 1, indicating the coexistence of multiple pinning mechanisms. 
While this shoulder structure persists in the 800 \hspace{1mm}$^{\circ}$C annealed sample, the surface pinning model again describes the experimental data (Fig.\hspace{1mm}\ref{fig9}(e)). 
The predominant pinning mechanism in Ta$_{1/6}$Nb$_{2/6}$Hf$_{1/6}$Zr$_{1/6}$Ti$_{1/6}$ annealed at 550 \hspace{1mm}$^{\circ}$C is known to be normal point pinning\cite{Kim:JMST2024}. 
In contrast, NbScTiZr annealed at 400 \hspace{1mm}$^{\circ}$C demonstrates considerable deviations from both the point and surface pinning models\cite{Kitagawa:EPJB2025}. 
In the 500 \hspace{1mm}$^{\circ}$C annealed Hf$_{10}$Nb$_{25}$Sc$_{25}$Ti$_{20}$Zr$_{20}$ sample, the pinning force at low fields is primarily governed by the point pinning mechanism, consistent with the low-field $J_\mathrm{c}$ behavior (Figs.\hspace{1mm}\ref{fig8}(a) and (b)) resembling that of Ta$_{1/6}$Nb$_{2/6}$Hf$_{1/6}$Zr$_{1/6}$Ti$_{1/6}$\cite{Kim:JMST2024}.
As illustrated in Figs.\hspace{1mm}\ref{fig9}(f)-(j), the normal surface pinning model predominantly accounts for the pinning behavior in most Hf$_{5}$Nb$_{45}$Sc$_{20}$Ti$_{15}$Zr$_{15}$ samples. 
The exception is the 500 \hspace{1mm}$^{\circ}$C annealed sample, whose experimental data lie between the surface and point pinning models.
This suggests that lattice strain, considered a point pinning source, partially contributes to the pinning mechanism. 
For the as-cast Hf$_{5}$Nb$_{45}$Sc$_{10}$Ti$_{5}$Zr$_{35}$ sample, a single pinning model fails to describe the data; instead, the surface pinning model agrees with low-field behavior, while the point pinning model captures the high-field response (Fig.\hspace{1mm}\ref{fig9}(k)). 
Neither model sufficiently explains the fishtail effect observed in the 400 \hspace{1mm}$^{\circ}$C annealed sample (Fig.\hspace{1mm}\ref{fig9}(l)). 
The experimental plots of the 500 \hspace{1mm}$^{\circ}$C and 800 \hspace{1mm}$^{\circ}$C annealed samples correspond well with the theoretical curves for point and surface pinning mechanisms, respectively (Figs.\hspace{1mm}\ref{fig9}(m) and (o)). 
Although the 600 \hspace{1mm}$^{\circ}$C annealed sample largely follows the surface pinning model, its low-field data do not substantially deviate from the point pinning model.

The pinning force analysis reveals that the normal point pinning mechanism is dominant over the surface pinning mechanism, particularly in the 500 \hspace{1mm}$^{\circ}$C annealed samples of Hf$_{10}$Nb$_{25}$Sc$_{25}$Ti$_{20}$Zr$_{20}$ and Hf$_{5}$Nb$_{45}$Sc$_{10}$Ti$_{5}$Zr$_{35}$. 
These samples exhibit increased lattice strain, as indicated by the reduced lattice parameters, similar to the 550 \hspace{1mm}$^{\circ}$C annealed Ta$_{1/6}$Nb$_{2/6}$Hf$_{1/6}$Zr$_{1/6}$Ti$_{1/6}$, which also follows the point pinning mechanism. 
Furthermore, in the 500 \hspace{1mm}$^{\circ}$C annealed Hf$_{5}$Nb$_{45}$Sc$_{10}$Ti$_{5}$Zr$_{35}$ sample, the phase instability, as evidenced by specific heat anomaly, may also contribute to the point pinning behavior.

\subsection{Enhanced $T_\mathrm{c}$ in Hf-Nb-Sc-Ti-Zr}

\begin{figure}
\begin{center}
\includegraphics[width=0.7\linewidth]{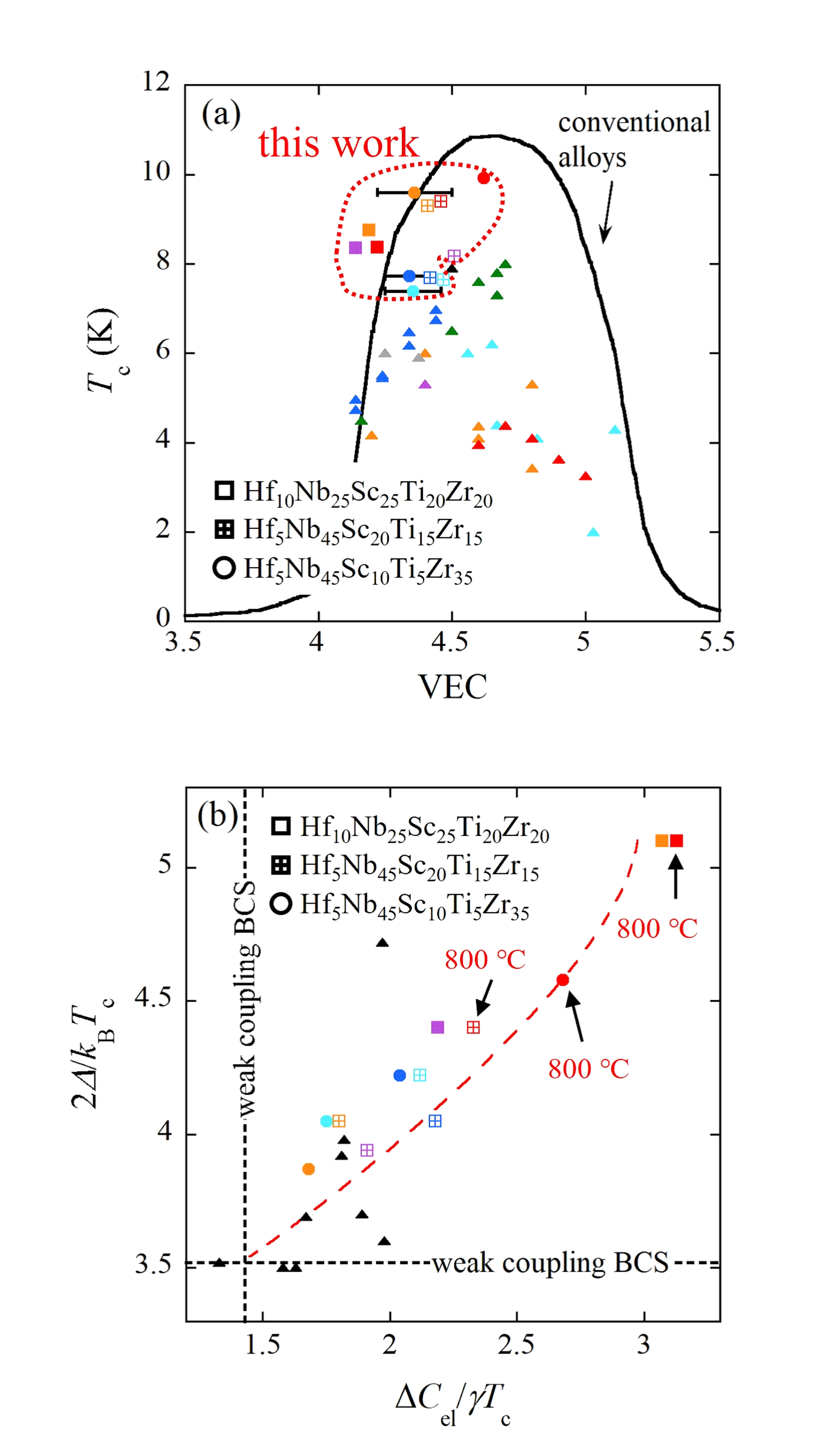}
\caption{\label{fig10}(a) VEC dependence of $T_\mathrm{c}$ for Hf-Nb-Sc-Ti-Zr HEAs. Within each Hf-Nb-Sc-Ti-Zr system, blue, light blue, purple, orange, and red represent the as-cast, 400 \hspace{1mm}$^{\circ}$C, 500 \hspace{1mm}$^{\circ}$C, 600 \hspace{1mm}$^{\circ}$C, and 800 \hspace{1mm}$^{\circ}$C samples, respectively. Data for typical quinary bcc HEA superconductors are also depicted by filled color triangles. The color correspondence for typical bcc HEAs is as follows: green-non-equimolar Hf-Nb-Ta-Ti-Zr\cite{Rohr:PNAS2016}; black-Ta$_{1/6}$Nb$_{2/6}$Hf$_{1/6}$Zr$_{1/6}$Ti$_{1/6}$\cite{Kim:AM2020}; blue-Al-Nb-Ti-V-Zr\cite{Harayama:JSNM2021}; purple-Hf$_{21}$Nb$_{25}$Ti$_{15}$V$_{15}$Zr$_{24}$\cite{Ishizu:RINP2019}; orange-HfNbTaTiZr, HfNbReTiZr, HfNbTaTiV, HfMoNbTiZr, TiHfNbTaMo, and ScVTiHfNb\cite{Jangid:APL2024,Marik:JALCOM2018,Sarkar:IM2022,Zeng:AQT2023,Kitagawa:JALCOM2022,Vrtnik:JALCOM2017}; light blue-Nb-Ta-Mo-Hf-W, Ti-Zr-Nb-Ta-W, and Ti-Zr-Nb-Ta-V\cite{Sobota:PRB2022,Shu:APL2022}; red-Ti-Hf-Nb-Ta-Re\cite{Hattori:JAMS2023}; and grey-Ti$_{0.5}$(ZrNbHfTa)$_{0.5}$ and Ti$_{0.5}$(VNbHfTa)$_{0.5}$\cite{Sobota:AM2025}. The solid line represents the Matthias rule of conventional binary or ternary transition metal alloys. (b) Correlation between $\frac{2\Delta(0)}{k_\mathrm{B}T_\mathrm{c}}$ and $\frac{\Delta C_\mathrm{el}}{\gamma T_\mathrm{c}}$ for Hf-Nb-Sc-Ti-Zr and several quinary HEAs. The red dashed line denotes the theoretical prediction calculated using Eqs.\hspace{1mm}(\ref{eq:strong1}) and (\ref{eq:strong2}).}
\end{center}
\end{figure}

Figure \ref{fig10}(a) presents the plots of $T_\mathrm{c}$ vs. VEC for the investigated Hf-Nb-Sc-Ti-Zr HEAs. 
$T_\mathrm{c}$ corresponds to the midpoint of the superconducting transition observed in $C_\mathrm{p}$($T$). 
The data for the as-cast and 400 \hspace{1mm}$^{\circ}$C annealed samples of Hf$_{10}$Nb$_{25}$Sc$_{25}$Ti$_{20}$Zr$_{20}$, as well as the 500 \hspace{1mm}$^{\circ}$C annealed sample of Hf$_{5}$Nb$_{45}$Sc$_{10}$Ti$_{5}$Zr$_{35}$, were excluded due to their extremely broad superconducting transitions. 
Error bars are included for the as-cast, 400 \hspace{1mm}$^{\circ}$C annealed, and 600 \hspace{1mm}$^{\circ}$C annealed samples of Hf$_{5}$Nb$_{45}$Sc$_{10}$Ti$_{5}$Zr$_{35}$, since both phase-separated bcc phases exhibit superconductivity in the as-cast and 400 \hspace{1mm}$^{\circ}$C annealed states and the chemical composition of the bcc phase in the eutectic structure cannot be accurately determined for the 600 \hspace{1mm}$^{\circ}$C annealed sample.
The plot also includes data for typical quinary HEAs, represented by colored triangles. 
Green, black, blue, and purple triangles indicate non-equimolar Hf-Nb-Ta-Ti-Zr, Ta$_{1/6}$Nb$_{2/6}$Hf$_{1/6}$Zr$_{1/6}$Ti$_{1/6}$, Al-Nb-Ti-V-Zr, and Hf$_{21}$Nb$_{25}$Ti$_{15}$V$_{15}$Zr$_{24}$, respectively\cite{Kim:AM2020,Rohr:PNAS2016,Ishizu:RINP2019,Harayama:JSNM2021}. 
Orange triangles represent data for HfNbTaTiZr, HfNbReTiZr, HfNbTaTiV, HfMoNbTiZr, HfMoNbTaTi, and ScVTiHfNb\cite{Jangid:APL2024,Marik:JALCOM2018,Sarkar:IM2022,Zeng:AQT2023,Kitagawa:JALCOM2022,Vrtnik:JALCOM2017}. 
Additional systems are indicated by light blue (Nb-Ta-Mo-Hf-W, Ti-Zr-Nb-Ta-W, and Ti-Zr-Nb-Ta-V), red (Ti-Hf-Nb-Ta-Re), and grey (Ti$_{0.5}$(ZrNbHfTa)$_{0.5}$ and Ti$_{0.5}$(VNbHfTa)$_{0.5}$) triangles\cite{Hattori:JAMS2023,Sobota:AM2025,Sobota:PRB2022,Shu:APL2022}. 
The solid curve represents the Matthias rule for conventional binary or ternary transition-metal alloys. 
Notably, at a given VEC, most Hf-Nb-Sc-Ti-Zr samples exhibit $T_\mathrm{c}$ values higher than those of typical quinary HEAs.
It is worth noting that the $T_\mathrm{c}$ of Ta$_{1/6}$Nb$_{2/6}$Hf$_{1/6}$Zr$_{1/6}$Ti$_{1/6}$ is comparable to that of Hf-Nb-Sc-Ti-Zr samples with no eutectic structure (i.e., the as-cast and 400 \hspace{1mm}$^{\circ}$C annealed Hf$_{5}$Nb$_{45}$Sc$_{10}$Ti$_{5}$Zr$_{35}$) or a partial eutectic region (as observed in the as-cast, 400 \hspace{1mm}$^{\circ}$C, and 500 \hspace{1mm}$^{\circ}$C annealed Hf$_{5}$Nb$_{45}$Sc$_{20}$Ti$_{15}$Zr$_{15}$). 
According to BCS theory, $T_\mathrm{c}$ depends on $\gamma$ and $\theta_\mathrm{D}$. 
For instance, the $\gamma$ and $\theta_\mathrm{D}$ values of as-cast Hf$_{5}$Nb$_{45}$Sc$_{10}$Ti$_{5}$Zr$_{35}$ are 7.62 mJ/mol$\cdot$K$^{2}$ and 207 K, respectively—comparable to 7.42 mJ/mol$\cdot$K$^{2}$ and 194 K in Ta$_{1/6}$Nb$_{2/6}$Hf$_{1/6}$Zr$_{1/6}$Ti$_{1/6}$\cite{Kim:AM2020}. 
This similarity further supports the comparable $T_\mathrm{c}$ between these two systems. 
In each Hf-Nb-Sc-Ti-Zr alloy, higher annealing temperatures lead to an expansion of the eutectic region, eventually encompassing the entire sample, enhancing $T_\mathrm{c}$, and establishing an another trend in the $T_\mathrm{c}$-VEC relationship.

To further investigate the underlying factors contributing to the enhanced $T_\mathrm{c}$ in Hf-Nb-Sc-Ti-Zr alloys, the relationship between $\frac{2\Delta(0)}{k_\mathrm{B}T_\mathrm{c}}$ and $\frac{\Delta C_\mathrm{el}}{\gamma T_\mathrm{c}}$ was examined for both Hf-Nb-Sc-Ti-Zr systems and selected quinary HEAs, as depicted in Fig.\hspace{1mm}\ref{fig10}(b). 
These ratios are not shown for the as-cast and 400 \hspace{1mm}$^{\circ}$C annealed Hf$_{10}$Nb$_{25}$Sc$_{25}$Ti$_{20}$Zr$_{20}$ samples and the 500 \hspace{1mm}$^{\circ}$C annealed Hf$_{5}$Nb$_{45}$Sc$_{10}$Ti$_{5}$Zr$_{35}$ sample due to their broad $C_\mathrm{p}$ transitions, from which reliable parameters could not be extracted.
The primary superconducting parameters of selected quinary HEAs are summarized in Table \ref{tab4}. 
Many HEAs fall within the strong-coupling regime, implying that strong coupling might be an intrinsic feature of bcc-structured HEAs. 
The parameters $\frac{2\Delta(0)}{k_\mathrm{B}T_\mathrm{c}}$ and $\frac{\Delta C_\mathrm{el}}{\gamma T_\mathrm{c}}$ in the strong-coupling limit are given by the following expressions\cite{Mitrovic:PRB1984,Marsiglio:PRB1986}:
\begin{equation}
\frac{2\Delta(0)}{k_\mathrm{B}T_\mathrm{c}}=3.53\left[1+12.5\left(\frac{k_\mathrm{B}T_\mathrm{c}}{\omega_\mathrm{ln}}\right)^{2}\ln\left(\frac{\omega_\mathrm{ln}}{2k_\mathrm{B}T_\mathrm{c}}\right)\right]
\label{eq:strong1}
\end{equation}
\begin{equation}
\frac{\Delta C_\mathrm{el}}{\gamma T_\mathrm{c}}=1.43\left[1+53\left(\frac{k_\mathrm{B}T_\mathrm{c}}{\omega_\mathrm{ln}}\right)^{2}\ln\left(\frac{\omega_\mathrm{ln}}{3k_\mathrm{B}T_\mathrm{c}}\right)\right],
\label{eq:strong2}
\end{equation}
where $\omega_\mathrm{ln}$ is the logarithmic average of phonon frequencies. 
The red dashed line in Fig.\hspace{1mm}\ref{fig10}(b) represents the theoretical relationship derived from these equations. 
The data points for Hf-Nb-Sc-Ti-Zr alloys align with this line, confirming that these systems are indeed strong-coupled superconductors. 
Most Hf-Nb-Sc-Ti-Zr samples annealed at lower temperatures exhibit slightly stronger coupling than other quinary HEA systems except Ta$_{1/6}$Nb$_{2/6}$Hf$_{1/6}$Zr$_{1/6}$Ti$_{1/6}$. 
This observation suggests that the specific elemental combination in Hf-Nb-Sc-Ti-Zr alloys contributes to the slightly elevated $T_\mathrm{c}$ values.
Higher annealing temperatures in Hf-Nb-Sc-Ti-Zr alloys promote the development of a fully eutectic microstructure, most prominently observed in the 800 \hspace{1mm}$^{\circ}$C annealed samples across all compositions. 
Figure \ref{fig10}(b) indicates that these 800 \hspace{1mm}$^{\circ}$C annealed samples shift the $\frac{2\Delta(0)}{k_\mathrm{B}T_\mathrm{c}}$ vs. $\frac{\Delta C_\mathrm{el}}{\gamma T_\mathrm{c}}$ relationship toward a stronger coupling regime, which correlates with the observed enhancement in $T_\mathrm{c}$. 
These analyses underscore the pivotal role of eutectic structures in enhancing superconducting transition temperature.

\begin{table}
\caption{$T_\mathrm{c}$, VEC, $\frac{2\Delta(0)}{k_\mathrm{B}T_\mathrm{c}}$, and $\frac{\Delta C_\mathrm{el}}{\gamma T_\mathrm{c}}$ of typical quinary HEAs.}\label{tab4}%
\begin{tabular}{cccccc}
\hline
HEA & $T_\mathrm{c}$ (K) & VEC & $\frac{2\Delta(0)}{k_\mathrm{B}T_\mathrm{c}}$ & $\frac{\Delta C_\mathrm{el}}{\gamma T_\mathrm{c}}$ & reference \\
\hline
Ta$_{34}$Nb$_{33}$Hf$_{8}$Zr$_{14}$Ti$_{11}$ & 7.3 & 4.67 & 3.5 & 1.63 & \cite{Kozelj:PRL2014} \\
(TaNb)$_{0.67}$(ZrHfTi)$_{0.33}$ & 7.8 & 4.67 & 3.7 & 1.89 & \cite{Rohr:PNAS2016} \\
(TaNb)$_{0.16}$(ZrHfTi)$_{0.84}$ & 4.5 & 4.16 & 3.6 & 1.98 & \cite{Rohr:PNAS2016} \\
Ta$_{1/6}$Nb$_{2/6}$Hf$_{1/6}$Zr$_{1/6}$Ti$_{1/6}$ & 7.9 & 4.5 & 4.72 & 1.97 & \cite{Kim:AM2020} \\
HfNbTaTiV & 5.0 & 4.6 & 3.69 & 1.67 & \cite{Sarkar:IM2022} \\
HfNbReTiZr & 5.3 & 4.8 & 3.92 & 1.81 & \cite{Marik:JALCOM2018} \\
HfMoNbTiZr & 4.1 & 4.6 & 3.5 & 1.58 & \cite{Kitagawa:JALCOM2022} \\
HfMoNbTaTi & 3.42 & 4.8 & 3.98 & 1.82 & \cite{Zeng:AQT2023} \\
HfNbScTiV & 4.17 & 4.2 & 3.52 & 1.33 & \cite{Jangid:APL2024} \\
\hline
\end{tabular}
\end{table}

Here, we examine the mechanism underlying strong-coupling superconductivity in Hf-Nb-Sc-Ti-Zr alloys. 
Using Eqs.\hspace{1mm}(\ref{eq:strong1}) and (\ref{eq:strong2}), the $\omega_\mathrm{ln}$ value of each sample was extracted, as summarized in Table \ref{tab3}. 
The degree of strong coupling was evaluated through the electron–phonon coupling constant $\lambda_\mathrm{ep}$, obtained from the McMillan equation modified by Allen and Dynes\cite{Allen:PRB1975}, expressed as
\begin{equation}
T_\mathrm{c}=\frac{\omega_\mathrm{ln}}{1.20}\exp\left(-1.04\frac{1+\lambda_\mathrm{ep}}{\lambda_\mathrm{ep}-\mu^{*}(1+0.62\lambda_\mathrm{ep})} \right).
\label{eq:AD}
\end{equation}
Assuming a constant Coulomb pseudopotential $\mu^{*}$=0.13, a value widely adopted for many transition-metal-based superconductors\cite{McMillan:PR1968}, $\lambda_\mathrm{ep}$ was calculated by substituting $T_\mathrm{c}$, $\omega_\mathrm{ln}$, and $\mu^{*}$ into Eq.\hspace{1mm}(\ref{eq:AD}). 
The resulting $\lambda_\mathrm{ep}$ values are listed in Table \ref{tab3}. 
Niobium is known as a strong-coupling superconductor with $\lambda_\mathrm{ep}$ $\sim$ 1.0\cite{Zeng:SCPMA:2023}, which is comparable to those of Hf$_{5}$Nb$_{45}$Sc$_{20}$Ti$_{15}$Zr$_{15}$ and Hf$_{5}$Nb$_{45}$Sc$_{10}$Ti$_{5}$Zr$_{35}$ at lower annealing temperatures. 
This indicates that the strong-coupling strength of Nb-rich HEAs mirrors that of elemental Nb.
The annealing-temperature dependence of $\lambda_\mathrm{ep}$ in Hf$_{5}$Nb$_{45}$Sc$_{20}$Ti$_{15}$Zr$_{15}$ and Hf$_{5}$Nb$_{45}$Sc$_{10}$Ti$_{5}$Zr$_{35}$ reveals that $\lambda_\mathrm{ep}$ decreases temporarily at intermediate temperatures (500-600 \hspace{1mm}$^{\circ}$C) but increases again at higher temperatures. 
Notably, the decrease in $\lambda_\mathrm{ep}$ within the 500-600 \hspace{1mm}$^{\circ}$C range is attributed to the enhanced $\omega_\mathrm{ln}$, suggesting phonon hardening. 
In these alloys annealed at 500-600 \hspace{1mm}$^{\circ}$C, fine microstructures comprising bcc and hcp phases emerge. 
The interfaces between these phases act as barriers to dislocation motion due to the mismatch of mechanical and structural properties, thereby inducing high hardness\cite{Xiong:JMST2021,Zhuang:Entropy2018}. 
Such interface strengthening is consistent with the pronounced lattice strain observed in Fig.\hspace{1mm}\ref{fig1}(d) and likely contributes to phonon hardening, thereby explaining the enhanced $\omega_\mathrm{ln}$. 
In contrast, coarsened microstructures in high-temperature-annealed samples of all systems, including Hf$_{10}$Nb$_{25}$Sc$_{25}$Ti$_{20}$Zr$_{20}$, exhibit reduced interfacial areas, which relax interface strengthening and consequently lead to phonon softening. 
Therefore, Hf-Nb-Sc-Ti-Zr alloys subjected to high-temperature annealing tend to exhibit decreased $\omega_\mathrm{ln}$, thereby promoting an enhanced $\lambda_\mathrm{ep}$.
We note that NbScTiZr displays a reduction in hardness upon high-temperature annealing\cite{Kitagawa:MTC2024}, corroborating that coarsened microstructures induce phonon softening.
L. Zeng et al. systematically compiled the superconducting parameters of transition-metal-based superconductors in the strong-coupling regime\cite{Zeng:SCPMA:2023}. 
According to their analysis, phonon softening plays a pivotal role in enhancing $\lambda_\mathrm{ep}$. 
This is consistent with the McMillan formalism, in which $\lambda_\mathrm{ep}$ is defined as $\lambda_\mathrm{ep}=\frac{N(E_\mathrm{F})\langle I^{2}\rangle}{M_\mathrm{w}\langle\omega^{2}\rangle}$, where $N(E_\mathrm{F})$ is the density of states at the Fermi level, $M_\mathrm{w}$ is the molecular weight, and $\langle I^{2}\rangle$ and $\langle\omega^{2}\rangle$ represent the averaged square of electron-phonon matrix element and the phonon frequency, respectively\cite{McMillan:PR1968}. 
In this framework, phonon softening corresponds to a reduction in $\langle\omega^{2}\rangle$, which enhances $\lambda_\mathrm{ep}$. 
We therefore infer that significant lattice strain is detrimental to strong-coupling superconductivity, whereas pronounced phonon softening, facilitated by the relaxation of interface strengthening during high-temperature annealing, ensures the manifestation of strong-coupling superconductivity in Hf-Nb-Sc-Ti-Zr alloys.

\section{Summary}
We have investigated the superconducting properties of Hf-Nb-Sc-Ti-Zr quinary eutectic HEAs subjected to heat treatments up to 800 \hspace{1mm}$^{\circ}$C, aiming to examine the universality of the deviation in the $T_\mathrm{c}$ vs. VEC relationship from the trend observed in typical quinary bcc HEAs. 
Additionally, we explored the correlation between microstructure and $J_\mathrm{c}$. 
The nominal compositions—Hf$_{10}$Nb$_{25}$Sc$_{25}$Ti$_{20}$Zr$_{20}$, Hf$_{5}$Nb$_{45}$Sc$_{20}$Ti$_{15}$Zr$_{15}$, and Hf$_{5}$Nb$_{45}$Sc$_{10}$Ti$_{5}$Zr$_{35}$—span a broad range of VEC values associated with the bcc phase.
The XRD patterns revealed a reduction in lattice parameters of the bcc phases upon annealing at 500-600 \hspace{1mm}$^{\circ}$C across all systems, indicative of significant lattice strain. 
In Hf$_{10}$Nb$_{25}$Sc$_{25}$Ti$_{20}$Zr$_{20}$ and Hf$_{5}$Nb$_{45}$Sc$_{20}$Ti$_{15}$Zr$_{15}$, samples annealed at lower temperatures exhibit partial eutectic regions, whereas annealing above 600 \hspace{1mm}$^{\circ}$C leads to an expanded eutectic region encompassing the entire sample. 
Conversely, the as-cast and 400 \hspace{1mm}$^{\circ}$C annealed Hf$_{5}$Nb$_{45}$Sc$_{10}$Ti$_{5}$Zr$_{35}$ samples consist solely of bcc phases, transforming into eutectic structures comprising bcc and hcp phases with increasing annealing temperature.
In all systems, $T_\mathrm{c}$ increases sharply from the 400 \hspace{1mm}$^{\circ}$C to the 600 \hspace{1mm}$^{\circ}$C annealed states, reaching a maximum of 9.93 K in the 800 \hspace{1mm}$^{\circ}$C annealed Hf$_{5}$Nb$_{45}$Sc$_{10}$Ti$_{5}$Zr$_{35}$ sample. 
Specific heat analyses reveal that nearly all samples exhibit characteristics of strong-coupled superconductors. 
The Hf$_{5}$Nb$_{45}$Sc$_{10}$Ti$_{5}$Zr$_{35}$ sample annealed at 500 \hspace{1mm}$^{\circ}$C demonstrates $J_\mathrm{c}$ exceeding the practical threshold of 10$^{5}$ A/cm$^{2}$, sustaining this level up to approximately 4 T at 4.2 K and 6 T at 2 K, indicating that this sample belongs to the highest-performing class of HEAs in terms of $J_\mathrm{c}$.
The pronounced lattice strain and inherent phase instability likely contribute to the elevated $J_\mathrm{c}$. 
Compared to representative quinary bcc HEAs, the $T_\mathrm{c}$ versus VEC plot for the Hf-Nb-Sc-Ti-Zr system underscores the enhanced $T_\mathrm{c}$ and delineates a distinct VEC dependence that deviates from typical bcc HEA trends. 
The relationship between $\frac{2\Delta(0)}{k_\mathrm{B}T_\mathrm{c}}$ and $\frac{\Delta C_\mathrm{el}}{\gamma T_\mathrm{c}}$ was evaluated to elucidate this result. 
The Hf-Nb-Sc-Ti-Zr samples reside in the strong-coupling regime relative to other quinary bcc HEAs. 
The expansion of the eutectic region in these samples further shifts this relationship toward the stronger-coupling domain, suggesting that the eutectic structure plays a pivotal role in enhancing $T_\mathrm{c}$.

\section*{CRediT authorship contribution statement}
Issei Kubo: Investigation. Yuto Watanabe: Investigation, Writing - reviewing \& editing. Shuma Kawashima: Investigation. Tomohiro Miyaji: Investigation. Yoshikazu Mizuguchi: Investigation, Formal analysis, Writing - reviewing \& editing. Terukazu Nishizaki: Investigation, Formal analysis, Writing - reviewing \& editing. Jiro Kitagawa: Supervision, Formal analysis, Writing - original draft, Writing - reviewing \& editing.

\section*{Acknowledgments}
J.K. is grateful for the support provided by the Comprehensive Research Organization of Fukuoka Institute of Technology and a Grant-in-Aid for Scientific Research (KAKENHI) (Grant No. 23K04570). Y.M. acknowledges the support from a Grant-in-Aid for Scientific Research (KAKENHI) (Grant No. 21H00151) and JST-ERATO(Grant No. JPMJER2201). T.N. acknowledges the support from a Grant-in-Aid for Scientific Research (KAKENHI) (Grant No. 24K08236) and the Takahashi Industrial and Economic Research Foundation. 

\section*{Declaration of Competing Interest}
The authors have no conflicts to disclose.

\section*{Data Availability Statement}
The data that support the findings of this study are available on reasonable request.

\end{document}